\documentclass[fleqn,usenatbib]{mnras}
\usepackage{newtxtext,newtxmath}
\usepackage[T1]{fontenc}
\usepackage{cancel}
\usepackage{fix-cm}
\usepackage{graphicx}
\usepackage{amsmath}
\usepackage{bm}
\usepackage{xr-hyper}
\usepackage{orcidlink}
\usepackage[dvipsnames]{xcolor}

\usepackage{bm,upgreek}
\usepackage{lineno}
\usepackage{soul}
\usepackage[normalem]{ulem}
\usepackage{caption}
\usepackage{overpic}
\newcommand{\appendixhead}%
{\begin{center}\textbf{\\Appendices\vspace{-0.5cm}}\end{center}}

\DeclareRobustCommand{\okina}{%
\raisebox{\dimexpr\fontcharht\font`A-\height}{%
    \scalebox{0.8}{`}%
  }%
}

\colorlet{mylinkcolor}{ForestGreen}
\colorlet{mycitecolor}{Red}
\colorlet{myurlcolor}{violet}

\allowdisplaybreaks

\usepackage{tikz,xcolor,hyperref}

\definecolor{lime}{HTML}{A6CE39}
\DeclareRobustCommand{\orcidicon}{\hspace{-1mm}
	\begin{tikzpicture}
	\draw[lime, fill=lime] (0,0) 
	circle [radius=0.16] 
	node[white] {{\fontfamily{qag}\selectfont \tiny \,ID}};
	\draw[white, fill=white] (-0.0525,0.095) 
	circle [radius=0.007];
	\end{tikzpicture}
	\hspace{-3mm}
}

\foreach \x in {A, ..., Z}{\expandafter\xdef\csname orcid\x\endcsname{\noexpand\href{https://orcid.org/\csname orcidauthor\x\endcsname}
			{\noexpand\orcidicon}}
}

\begin{document}
\title{Modelling redshift-space distortions with tracer-centred split densities}
\author[A. Gon et al.]{\newauthor
Aritra K. Gon$^1$\thanks{E-mail: agon@roe.ac.uk}\orcidlink{0000-0002-0004-9563}, Yan-Chuan Cai$^1$, Mark Neyrinck$^{2,3}$\orcidlink{0000-0002-2618-5790}, Andrew Repp$^4$\orcidlink{0000-0002-0116-1039}, Ravi K. Sheth$^5$\orcidlink{0000-0002-2330-0917}, István Szapudi$^2$
\\\\
$^1$Institute for Astronomy, University of Edinburgh, Royal Observatory, Blackford Hill, Edinburgh, EH9 3HJ, UK\\
$^{2}$Institute for Astronomy, University of Hawai\`i at M\=anoa, 2680 Woodlawn Drive, Honolulu, HI 96822, USA\\
$^{3}$Blue Marble Space, 600 1st Avenue, 1st Floor, Seattle, Washington 981049, USA\\
$^{4}$Christian Liberty Academy, 16-675 Milo St., Kea{\okina}au, HI 96749, USA\\
$^{5}$Center for Particle Cosmology, University of Pennsylvania, 209 S 33rd Street, Philadelphia, PA 19104, USA
}
\date{\\}           

\maketitle           
\begin{abstract}
Analysing galaxy clustering in split environmental densities has been demonstrated to offer an advantage over standard two-point clustering for extracting cosmological information. A major challenge to realising this potential with the density-split (DS) technique is that the observations are in redshift space.  This complicates its modelling, primarily because the density-split centres may not be conserved between real and redshift space. We develop a modelling framework that splits on the tracer-centred density PDF, rather than the volume-weighted density PDF commonly used in previous studies. We derive the linear bias associated with the threshold density in redshift space, enabling us to connect large-scale DS clustering to the density PDF. We find that the threshold density field consists of doughnut-like structures, which we term `Island(s)'. We propose an empirical model, analogous to the halo model framework, that comprises linear, non-linear, and stochastic terms and accurately reproduces the anisotropic redshift-space clustering of tracers in each DS bin. Using this model, we confirm that DS clustering can provide tighter constraints on cosmological parameters than the standard two-point statistic. Furthermore, we show that the gain in cosmological information from DS analyses relative to standard two-point statistics arises primarily from knowledge of the density PDF, which incorporates small-scale perturbation modes below the smoothing scale.
\end{abstract}
\section{introduction}
Galaxy redshift surveys and lensing surveys of the large-scale structure have enabled us to constrain models of the Universe and test theories of gravity by measuring matter perturbations traced by galaxies or gas. While field-level inferences are now being actively pursued \citep[e.g.][]{Shu_2008, Jasche_2010, Wandelt_2013, Wang_2014, Lavaux_2019, Schmidt_2019, Elsner_2020, Schmidt_2020, Kostic_2023, Andrews_2023, Bayer_2023, Doeser_2024, Schmidt_2025}, reduction of 3D datasets via summary statistics still serves as the major means of extracting cosmological information from data. A standard summary statistic of the large-scale structure is the two-point correlation function (2PCF), or its Fourier version, the power spectrum. These are known to be sufficient statistics for Gaussian fields, but non-linear evolution of the density field towards the late-time Universe 
causes the field to deviate from Gaussianity. 
This motivates thinking beyond two-point statistics in order to capture the cosmological information leaked out of two-point statistics \citep[e.g.][]{Peebles_1975ApJ, Fry_1978ApJ, Fry_1982ApJ, Neyrinck_2009ApJ, Neyrinck_2011ApJ, Wang_2011ApJ, Chiang2014, Schmittfull_2015PRD, White2016JCAP, Friedrich2018, Gruen2018, Nadathur_2019PRD, Hawken_2020JCAP, Enrique2021, Banerjee_2021MNRAS, Repp2022, Georgios_2022PRD,  Paillas2023, Azadeh_2024PRD,
Paillas2024MNRAS, Cheng_2025JCAP, Harscouet_2025OJAp, Francesco_2026JCAP,   Ebina_2026, Chand2026arxiv}. Many of these techniques have been discussed in "Beyond 2-pt Challenge” \citep{ThBeyond-2pt_2025}. The technique of density-split clustering is one of them. It has been shown to be effective in capturing more cosmological information than conventional two-point statistics \citep{Enrique2021,Giblin_2023MNRAS,Paillas2023}. 

The advantages of DS clustering can be appreciated through the summary of its analysis pipeline: 

(1) Evaluate the PDF of the density field at a smoothing scale, typically chosen at the quasi-nonlinear regime (e.g., on the order of 10 $h^{-1}\rm Mpc$) such that the density PDF is non-Gaussian -- a crucial condition for obtaining information beyond two-point statistics. 

(2) Split the density PDF into $M$ bins of low and high densities. This creates sub-volumes of the density field that satisfy the density thresholds. Their centres are called the DS centres. 

(3) Measure the cross-correlations between those DS centres in each density bin with the entire density field, as well as the auto-clustering of those DS centres. 

So, instead of having one global measurement of the 2PCF or power spectrum from the density field, we will have $M$ cross-correlations and $M$ auto-correlations in DS clustering. These cross-correlations are measurements of conditional clustering, conditioned on the environmental densities. The DS auto-correlations are the clustering of the density environments themselves. In addition, what is not so explicit is that the density PDF evaluated at the smoothing scale is naturally included as part of the data vector. It serves as a sampling of the non-Gaussian PDF at the smoothing scale. The characteristics of DS clustering are summarised as follows. \\
$\bullet$ It captures some information beyond the two-point clustering because the DS clustering samples the non-Gaussian density PDF at the smoothing scale. \\
$\bullet$ The velocity field in each DS bin is more Gaussian than the Global velocity field \citep[e.g.][]{Tinker2007, Enrique2021, Gon2025}. This eases the modelling of redshift-space distortions (RSD), making it easier to reach a higher accuracy compared to the Global 2PCF on the same scales \citep{Enrique2021}. \\
$\bullet$ Density-split separates converging flows versus diverging flows of the velocity field. This avoids the cancellations in the Global-clustering measurement when those flows of opposite sign are added together in the Global average -- effectively turning some dispersions (noise) into signal \citep{Gon2025}. This contributes to the gain of cosmological information over two-point statistics in the analysis of the pairwise kinetic Sunyaev-Zeldovich effect  \citep{roper2026arxiv}. 

In 2D, the modelling of DS statistics was developed in the context of weak lensing \citep{Friedrich2018}, followed by \citep{Burger2022A&A} for more general filters. In 3D, an analytical model was presented by \citet{Pinon_2025JCAP}, but in real space. There are, however, major challenges in the modelling of DS clustering in 3D redshift space. The main problem arises from the fact that the mapping between real and redshift space for the threshold densities may not be conserved \citep[e.g.][]{Nadathur2019, Correa_2021}. For a given density threshold, a voxel selected in real space may not necessarily be selected in redshift space, and vice versa. In addition, the centres of voxels shift between redshift and real space, and the shifts are contributed by peculiar velocities of the voxels, which are non-local. This complicates the modelling of RSD. Efforts have been made to overcome this challenge. One way is to move tracers (galaxies)\footnote{Note that we will be using sub-sampled particles from simulations as tracers of the density field for our analyses. In this simplest assumption, they are referred to as simulated galaxies, or simply galaxies, throughout the paper.} back to real space through reconstruction, and conduct DS-RSD analyses from the reconstructed galaxy field. With this, both the measurement and the modelling can start from the real-space DS centres, as it was done in void-RSD \citep[e.g.][]{Nadathur2019}. While this method works in principle, its accuracy depends on the accuracy of the reconstruction, which depends on the model and model parameters of the reconstruction.

To overcome the above challenge of modelling DS clustering, in this paper, we keep the spirit of DS clustering while exploring a different way of analysis in detail. The subtlety is to deviate from the conventional counts-in-cells (CIC) measurement of the density PDF, which is an estimate of the PDF averaged over the whole volume. Instead, we will evaluate the density PDF at the locations of tracers (galaxies or halos) only. In essence, we switch from volume-weighting to centring on tracers, as was done in the first density-split analysis of galaxy clustering \citep{Abbas2007}. By focusing on the densities around the tracers only, we naturally avoid the issue of conservation of voxels between real and redshift space. This allows us to model DS clustering in the same way as the global 2PCF, once the effects of the smoothing window function are accounted for.

Our paper is organised as follows. In Section \ref{sec:DS_bias_calc}, we lay down the new way of splitting the density and make connections to the previous method. In Section~\ref{sec:PDF}, we derive the linear bias associated with a threshold density in redshift space. 
In Section \ref{sec: DS_cluster_model}, we write down the model for the new DS clustering and compare it with the approximate lognormal simulations. Section \ref{sec:fisher_forecast} compares the cosmological constraints from the new DS clustering with those from two-point statistics. We draw our conclusions and discuss our results in Section \ref{sec:conclusion}.
\begin{figure}
\centering
\includegraphics[width=0.48\textwidth]{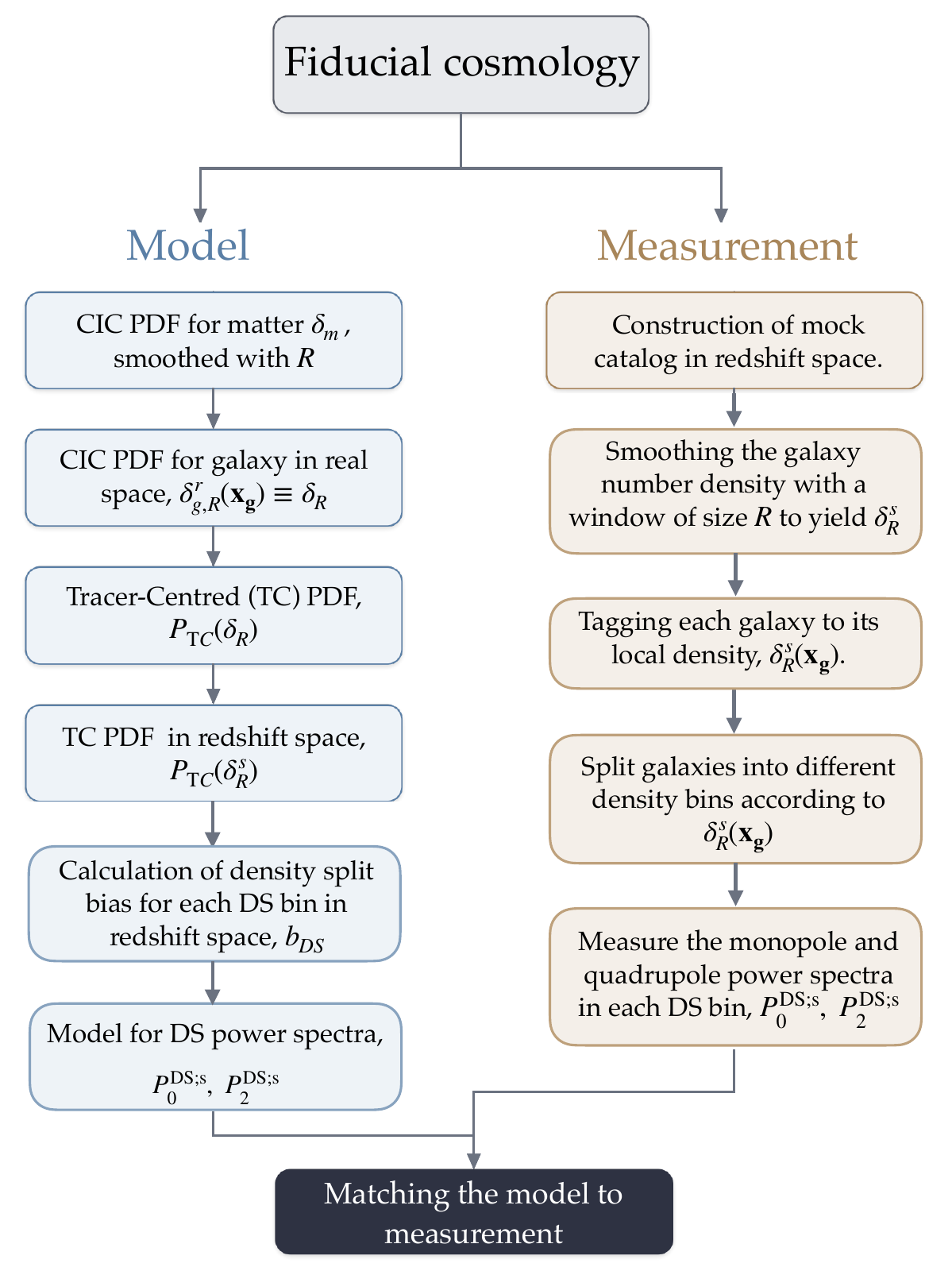}
\caption{The flowchart of the major steps of the modelling and the measurement of the density-split (DS) power spectra.}
\label{fig:flowchart}
\end{figure}

\section{Overview of modelling density-split clustering}
\label{sec:DS_bias_calc}
Our aim is to model the observed DS clustering in redshift space directly. 
We will evaluate the density PDF on locations of galaxies. Each galaxy will be tagged with an environmental density for its clustering analysis. So, our DS clustering is essentially a marked statistic. This also makes the modelling of clustering on different environmental densities the same as the global 2PCF, in principle. An overview of the method is summarised in a flowchart in Fig.~\ref{fig:flowchart}.
\subsection{An observational perspective}
In observations, the starting point is the redshift-space distribution of galaxies/halos. 
We denote real-space positions as $\bf x$ and redshift-space positions as $\bf s$.  Starting from the redshift-space positions of galaxies, we can create a field of galaxy number density contrast,   $\delta^s({\bf s})$ by putting the galaxies on a grid. We smooth the field with a window function $W({\bf s},R)$ to obtain the smoothed density field $\delta_R^s({\bf s})=\int W(|{\bf s-s' }|,R)\ \delta^s({\bf s'})\ d^3s'$. We then evaluate the smoothed densities at the locations of galaxies to obtain a PDF of the field. We call this the `tracer-centred' (TC) PDF, $P_\mathrm{TC}[\delta_R^s(\bf{s_g})]$. This is the PDF conditioned on the locations of the tracer galaxies; it is different from the usual counts-in-cells PDF, which we refer to here as the volume-weighted (VW) PDF, $P[\delta_R^s ({\bf s})]\equiv P(\delta_R^s)$. So, each galaxy is tagged with an environmental density $\delta_R^s(\bf{s_g})$. We split the galaxy sample into multiple environmental density bins and measure the power spectra of galaxies which fall into the same bin.

\begin{figure}
\includegraphics[width=0.476\textwidth]{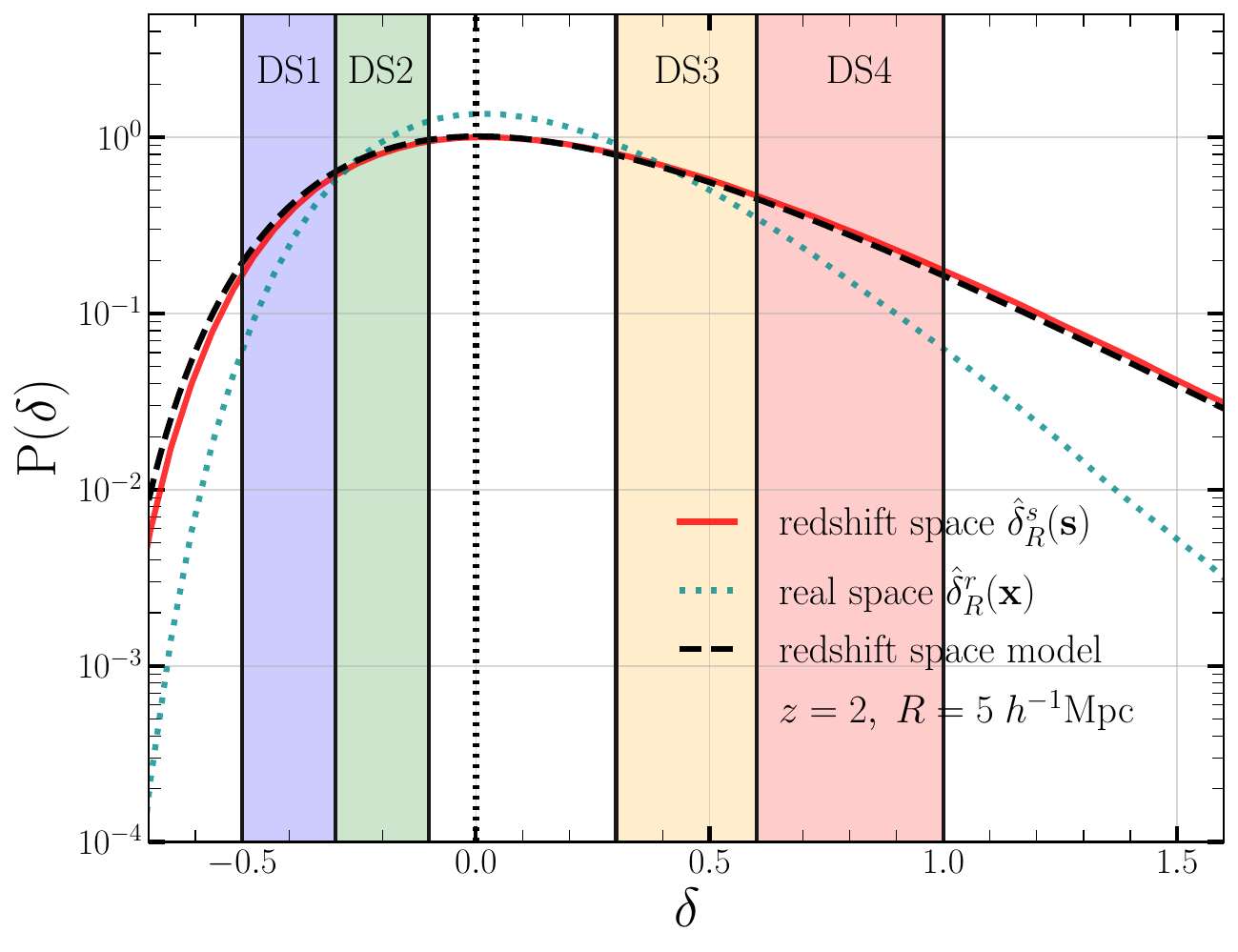}
\caption{The tracer-centred density PDFs measured from the approximate lognormal simulations at $z=2$. They are smoothed with a Gaussian filter of $R=5~h^{-1}$Mpc. The green and red curves are the real- and redshift-space versions, respectively. The vertical colour regions show the four density-split (DS) bins chosen for the measurements of the power spectra. DS1-4, as labelled in the legend, represent low to high environmental densities. The black dashed line is our model prediction, which agrees well with the measurement.} 
\label{fig:PDF_real_redshift_space}
\end{figure}
\subsection{Simulation setup}
\label{sec:simulation}
To provide benchmarks to test our model, we generate simulated 'galaxy' (a term we use to refer to the tracers) catalogues at redshift $z=2$ with {\sc mockfactory}\footnote{https://github.com/cosmodesi/mockfactory} package. We used a flat-$\Lambda$CDM cosmology ($\Omega_{\rm m} = 0.3175$, $\Omega_{\rm b} = 0.049$,
$h = 0.6711$, $n_{\rm s} = 0.9624$, $\sigma_8 = 0.834$, $M_\nu = 0.0\ {\rm eV}$, and $w = -1$) \citep{Planck_2018} evolved to $z=2$ using the Zeldovich approximation. We create the mock galaxy catalogue by randomly selecting dark matter particles such that the resulting number density is $6\times 10^{-2}\ h^{3}\rm Mpc^{-3}$ in cubic boxes of 1000 \ $h^{-1}\rm Mpc$ aside. In effect, these tracers have bias $b_g=1$. The velocity field of the simulated galaxies is linearly coupled to the original Gaussian density field, which was sourced from the linear matter power spectra at $z=2$. We use the velocities to displace the simulated galaxies into redshift space using the plane-parallel approximation. Due to the linear coupling, there is no Finger-of-God effect from velocity dispersions in the redshift-space galaxy clustering. We then smooth the redshift-space galaxy number density field with a Gaussian window of 5 $h^{-1}$Mpc and evaluate the smoothed density at the location of each galaxy. The smoothed densities provide tags of the environmental densities for the simulated galaxies, allowing us to split the galaxy sample into four density bins, as shown in Fig.~\ref{fig:PDF_real_redshift_space}. These density bins are DS1: $\hat\delta^s_R\in [-0.5,-0.3]$, DS2: $\hat\delta^s_R\in [-0.3,-0.1]$, DS3: $\hat\delta^s_R\in [0.3,0.6]$, and DS4: $\hat\delta^s_R\in [0.6,1.0]$. They respectively contain $7.5\%$, $16.0\%$, $19.1\%$, and $12.2\%$ of the total number of galaxies. 

\begin{figure}
\hspace{-0.1cm}
\includegraphics[width=0.485\textwidth]{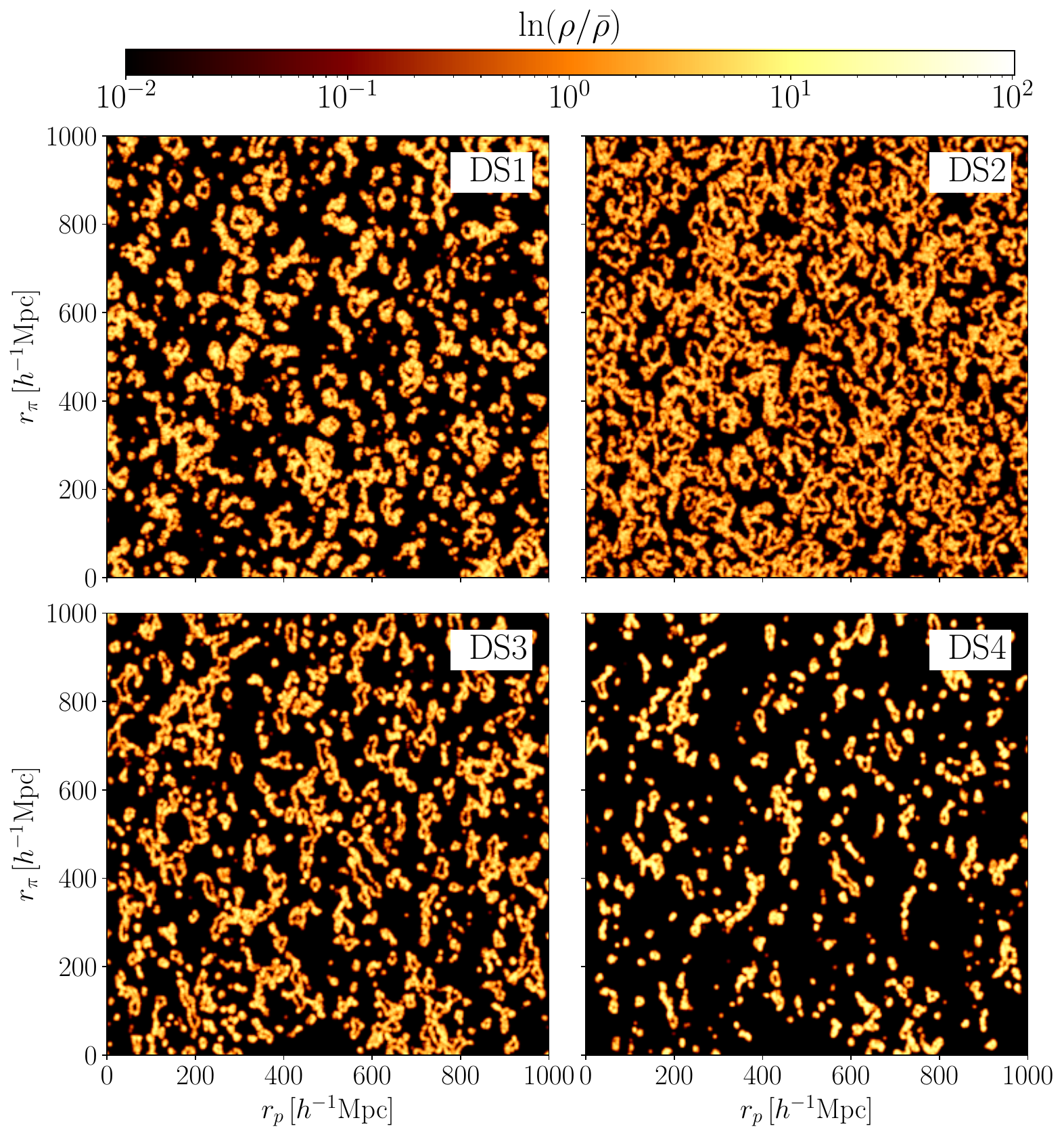}

\caption{2D number density fields for a slice with thickness $2\ h^{-1}$Mpc along the $z$-axis from a 3D box of 1$h^{-1}$Gpc on a side. Each panel shows one of the four DS bins defined in Fig.~\ref{fig:PDF_real_redshift_space}. We first select all simulated galaxies in a given density bin and construct the 3D density field on the mesh from the galaxy catalogue using the cloud-in-cell mass assignment scheme. Then we take a slice of the 3D density field along the $z$-axis and average the number density along that slice to obtain the 2D number density field. We refer to the doughnut-like structures with fuzzy boundaries as `Island(s)'.}
\label{fig:2d_number_density}
\end{figure}

These fractions do not sum to unity because we exclude the extreme edges and the region surrounding $\delta^s_R=0$. Density bins at the extreme edges of the PDF will be dominated by shot noise as the number density is small, and they contain little Fisher information \citep{Repp2025MNRAS}. So we stick to the regime where the PDF is at most an order of magnitude smaller than the maximum. The central regions around $\delta^s_R=0$ were excluded as the amplitude of clustering is small, and they also contain little Fisher information \citep{Repp2025MNRAS}.

After allocating the galaxies into the four density bins, we generate the four galaxy number density fields. For each of the four density bins, a 2D slice of the 3D galaxy number density field with a thickness of $2\ h^{-1}$Mpc along the $z$-axis from the 3D box of 1$h^{-1}$Gpc aside at $z=500 \,h^{-1}$Mpc, is shown in Fig.~\ref{fig:2d_number_density}. We can see the striking differences in the distribution of galaxies in these four density bins. The highest density bin (DS4) has sparsely segregated high-density peaks. There are also noticeable doughnut-like blobs of high concentrations of galaxies identified after the smoothing, which we refer to as `Islands'. These are due to imposing both upper and lower bounds on the density thresholds. The size, height and sparsity of those Islands vary among different density bins. 

We then measure the redshift-space power spectra for galaxies in these four density bins. These power spectra in different density bins follow linear theory but are complicated by the measurement window. They provide the benchmark for validating our model.

\subsection{Modelling framework}
In theory, we stay with the key idea of using the smoothed environmental densities as tags, or weights, for the galaxies. 
In analogy with the Halo Model, we propose an empirical model to describe the clustering of tracers, which are distributed in these Islands. On the largest scales, the clustering of Islands becomes important. The two-Island term stays linear, i.e. with linear redshift-space distortions and linear bias parameters, which are straightforward to write down. The key quantity for the two-Island term involves the computation of the large-scale linear biases associated with the environmental density bins in redshift space, which are referred to as the density-split biases, $b_{\rm DS}$. To compute these biases, we need a model for the tracer-centred density PDF $P_\mathrm{TC}[\delta_R^s(\bf{s_g})]$. For this, we start from the real-space, volume-weighted counts-in-cells PDF, $P[\delta_R(\bf{x})]$ \citep{Uhlemann2016,Uhlemann_2025}, and map it to the redshift-space tracer-centred density PDF. In linear theory, these biases are the key quantities needed to predict the amplitude of redshift-space DS clustering. We now specify this modelling process.  

\section{Density-split bias in redshift space}
\label{sec:PDF}
\subsection{Counts-in-cells}
The first step is to model the galaxy density PDF in real space. This is also called counts-in-cells \citep{Peebles1980, Hubble_1934ApJ, Lam2008, Uhlemann_2020MNRAS,Repp2021, Chudaykin_2023JCAP}, and the volume-weighted PDF in our terminology. 

In general, the smoothed galaxy fluctuation field differs from the smoothed matter field, and this is often called `bias'. But in our case this bias equals 1. Hence, in this limit, we just need a model for the PDF of the underlying matter density field smoothed at that same scale.
For this, we follow \citep{Uhlemann2016, Uhlemann2016a} to compute the PDF of the underlying matter density field $\delta_m({\bf x} )$ smoothed at scale $R$: 
\begin{align}
    P(\delta_{m,R}=\delta|\sigma)=&\sqrt{\left(\frac{\Psi''_R(\rho)+\Psi'_{R}(\rho)/\rho}{2\pi\sigma^2(R)}\right)}
    \exp\left(-\frac{\Psi_R(\rho)}{\sigma^2(R)}\right),
\label{dm_pdf}
\end{align}
where $\rho = 1+\delta$,
\begin{align}
    \Psi_R(\rho)=\frac{\sigma^2(R)}{2\sigma^2(R\rho^{1/3})}\tau^2(\rho),
    \qquad 
\end{align}
where $\tau=\nu(1+\rho^{-1/\nu})$ and $\nu=21/13$. This expression is derived for a top-hat filter of radius $R$, assuming that the variance, $\sigma^2(R)$, is small, and the density is rescaled such that $\langle \rho \rangle=1$. This was shown to be accurate down to the smoothing scale of $\sim 20~{h^{-1}}$Mpc at redshift zero. We are using a much higher redshift to reduce the non-linear effect, so we expect this analytic expression to hold for a much smaller smoothing scale.  We show later that this is indeed the case. Henceforth, to reduce clutter, we will denote $\delta_{g,R}$ as $\delta_R$. Any redshift-space quantities will be denoted by a superscript $s$ as necessary. 
\subsection{From random to tracer-centred counts}
We now convert the volume-weighted PDF for the real-space galaxy number densities into its tracer-centred version, $P_\mathrm{TC}$. (Recall that the volume-weighted PDF is referred to as $P$ with no subscript.) 
Operationally, the tracer-centred PDF is just 
$$P_\mathrm{TC}(N) = \frac{\sum_g \delta_K(N_{g,R} = N)}{\sum_g 1}$$
where $N_{g,R}$ is the number of galaxies in a cell of size $R$ centred on galaxy $g$, and the denominator is just the total number of galaxies.  We then define $\hat \delta_{g,R} \equiv N/\bar{n}_gV_{R} - 1$, where $V_R$ is the volume of the smoothing filter.
We show in the Appendix \ref{Appendix_TC_PDF} that

\begin{align}
    P_\mathrm{TC}(\hat \delta_{g,R}) &\equiv P(N|{\bf x_g})\nonumber\\&= \int d\delta_{g,R} \,P_{\rm Poisson}(N|\delta_{g,R},{\bf x_g})\, P(\delta_{g,R}|{\bf x_g})\nonumber\\
    &\approx \left(1+\frac{\bar{\xi}(R)}{\sigma^2(R)}\, \left(\hat\delta_R-\frac{1}{\bar N}\right)\right) P\left(\hat \delta_R-\frac{1}{\bar N}\right).
\end{align}

\begin{figure}
\includegraphics[width=0.48\textwidth]{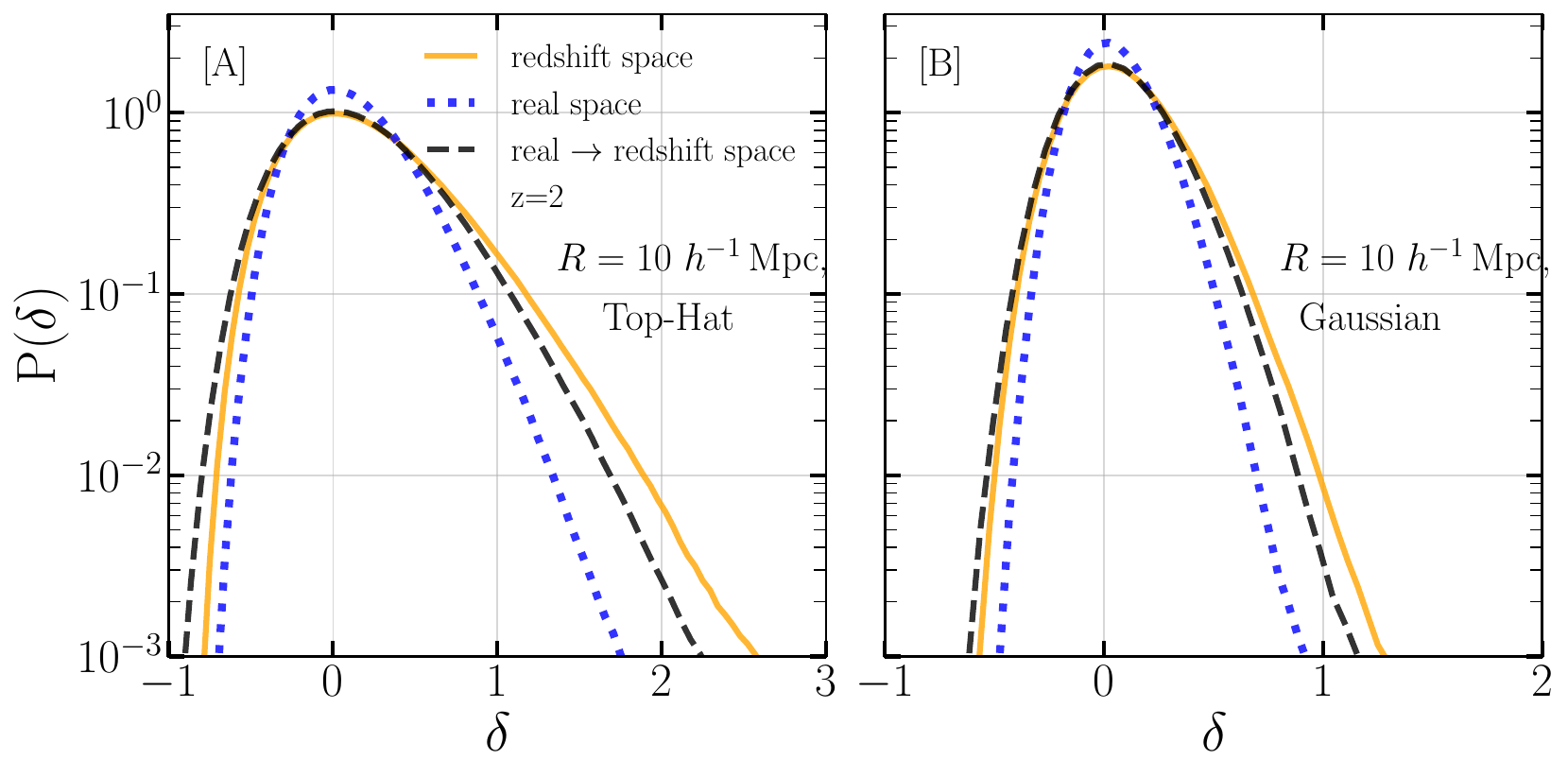}
\caption{Panels [A] and [B] show the distribution (PDF) of the simulated number density contrast in real space (in dotted blue) and redshift space (in solid orange) lines at a smoothing scale of $10\, h^{-1}{\rm Mpc}$ with two different types of filters, respectively. The mock galaxy catalogue is created using {\sc mockfactory} with a linear bias $b_g=1$, redshift $z=2$, and the smoothing is performed at the galaxy centres (tracer-centred). The dashed black curves show the redshift-space PDFs when they are remapped from real space using the formula provided in Eq.~\ref{real_to_rsd_space}. }
\label{fig:real_redshift_space_density_pdf}
\end{figure}
Eq.~\ref{dm_pdf} was designed and tested with a spherical top-hat filter. However, we use a Gaussian filter in the measurements which follow. We have checked that if we choose an equivalent top-hat smoothing scale such that $\sigma^2(R_G)=\sigma^2(R_{TH})$ (this makes $R_G\approx R_{TH}/\sqrt{5}$ ) the model still works well \citep{Chudaykin2026arXiv}. Now that we have the tracer-centred PDF in real space, the next step is to obtain the corresponding PDF in redshift space. 

\begin{figure*}
\centering
\includegraphics[width=0.995\textwidth]{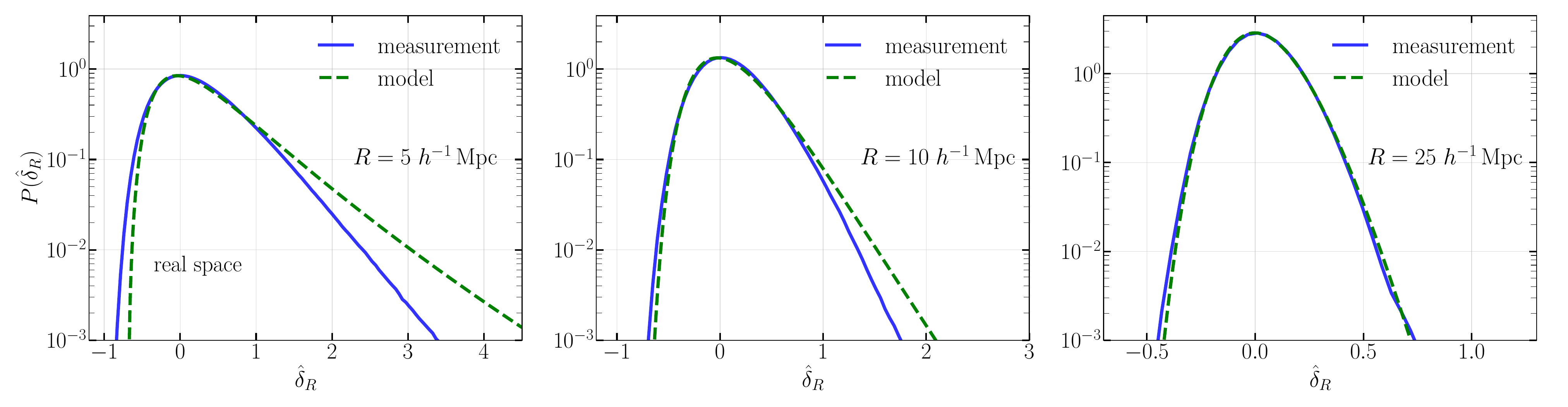}
\includegraphics[width=0.995\textwidth]{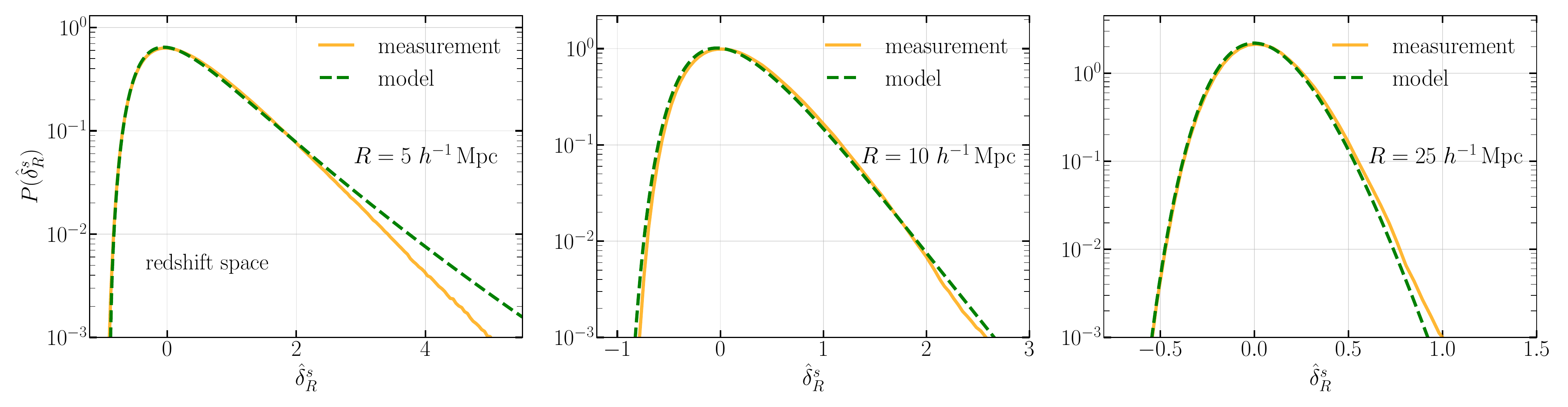}
\caption{Upper Panel: Comparing the real-space tracer-centred PDFs between the approximate lognormal simulations (solid) and the model predictions (dashed) with different smoothing scales of a top-hat filter at $z=2$.  Lower Panels: The same as the upper panels but in redshift space. The model works progressively better as the smoothing scale increases, effectively decreasing the variance of the field. Panels from left to right show the results from different smoothing scales, as labelled.}
\label{fig:model_measurement_pdf_comparison_same_z_diff_sm}
\end{figure*}
\subsection{From real space to redshift space}
A key feature of the tracer-centred density split is that the number of DS centres are now guaranteed to be the same in both real and redshift space, because we are tracking the exact same sub-sample of galaxies in real and redshift space. 
Each galaxy's smoothed density is mapped from $\hat\delta_R$ to $\hat\delta^s_R$, so the PDFs can be related directly.
All that is needed is a model for $\hat\delta^s_R(\hat\delta_R)$ at the smoothing scale $R$, which is linear on large scales, but deviates from linearity on small scales. Assuming linear theory, we have 
\begin{align}    
\hat\delta^s_R(\mu) = \hat\delta_R+ \frac{\beta_g}{3}{\bar{\hat\delta}_R}
    +\beta_g\mu^2
      \bigl[\hat\delta_R-{\bar{\hat\delta}_R}\bigr],
\end{align}
where $\beta_g=f/b_g$, $f$ is the growth rate, $\bar{\hat\delta}_R=(3/r^3)\int_0^r \hat\delta_R r'^2 dr' $, and $ b_g $ is the tracer linear bias, which, in our case, is set equal to 1 \citep{Cai2016MNRAS}. Averaging over all $\mu$'s yields

\begin{equation}
 \hat\delta^s_R \simeq(1+ \frac{1}{3}\beta_g)\,\hat\delta_R .
 \label{eq:deltas}
\end{equation}
Fig.~\ref{fig:scatter_plot} shows that this prescription works well, on average, for top-hat filters for smoothing scales $R \gtrsim 10~h^{-1}$Mpc, but there is noticeable scatter around the mean that increases as $R$ decreases. 
If we assume that Eq.~\ref{eq:deltas} is accurate, and the scatter around it can be ignored, then the real and redshift-space density PDFs are related by:
\begin{align}
    P_\mathrm{TC}^s(\hat\delta^s_R)=\frac{1}{\langle(1+\mu^2 \beta_g)\rangle }\,P_\mathrm{TC}^r\left(\frac{\hat\delta^s_R}{\langle(1+\mu^2 \beta_g)\rangle}\right).
    \label{real_to_rsd_space}
\end{align}
Fig.~\ref{fig:real_redshift_space_density_pdf} shows the real (dotted) and redshift-space (solid) PDFs measured in the $z=2$ outputs of the {\sc mockfactory} simulation, for top-hat and Gaussian filters of scale $R=10~h^{-1}$Mpc.
Dashed curves show the result of mapping the real-space PDF to redshift-space; it works reasonably well. We compare the model and the measured tracer-centred PDFs both in real and redshift space, for 3 different top-hat smoothing scales, in Fig.~\ref{fig:model_measurement_pdf_comparison_same_z_diff_sm}. For a given redshift, the model works progressively better as $\sigma$ decreases. For a fixed redshift, this amounts to increasing the smoothing scale $R$; Or, for a fixed $R$, it is equivalent to increasing redshift.

\subsection{Bias model for split densities}
Having modelled the non-Gaussian  PDF in redshift space, we can proceed to apply the linear model for density splits. We start by assuming the PDF obtained in Eq.~\ref{real_to_rsd_space} is approximately lognormal. Next, we define 
  $A = \ln(1 + \hat\delta_R)$, 
which should be approximately Gaussian. 
We tested these assumptions numerically and found that the approximation improves for larger smoothing scales, as expected.  

Following \citet{Repp2022}, we define the DS bias factor with respect to the biased tracer field as 
\begin{align}
\label{DS_bias}
    b_{\rm DS} = \sqrt{\frac{\sigma^2_A}{\sigma^2_\mathrm{TC}}}\left(\frac{A_{DS}-\langle A\rangle}{\sigma^2_A} \right),
\end{align}
where $\langle A\rangle =\int A P_\mathrm{TC}(A) \, dA$, $\sigma^2_A = \int (A-\langle A\rangle )^2P_\mathrm{TC}(A) \,d A$ and 
$\sigma^2_\mathrm{TC} = \langle (\hat\delta_R -\langle\hat\delta_R\rangle_\mathrm{TC})^2\rangle_\mathrm{TC}$.
Since $b_{\rm DS}$ is the bias with respect to the galaxy field, then, because $\delta_g\approx b_g\delta_m$, the bias with respect to the mass field is $b_{\rm DS}b_g$.  On large scales, the term under the square root is approximately unity, and the term in brackets is a Gaussian variate divided by its variance.  Hence, this agrees with the expression for the tracer-dependent density-split bias in \cite{Abbas2007}. The validity of the model and the comparison of the DS bias prediction with the measurements are made in Section \ref{sec: model_simulation_test}.

In summary, the above models take us forward from the real-space distribution of galaxies/halos step by step to the redshift-space linear biases of galaxies/halos in different density environments. These allow us to model redshift-space distortions with the linear assumption of biases in DS clustering and compare them with measurements in redshift space in simulations and observations. 
\section{Modelling density-split clustering}
\label{sec: DS_cluster_model}

Our goal here is to provide a backbone framework for DS clustering. The main idea is to demonstrate that it can capture the redshift-space clustering directly without any reconstruction or emulation, in contrast to what has been done in previous DS clustering analyses. In particular, motivated by the similarity (cf. previous section) of the clustering between the smoothed density field and a halo field, we create an analogous Halo Model-like framework for DS clustering. We intend to spell out a more `first-principles' description of this decomposition in a separate paper (Sheth et al., in preparation). For now, we simply capture the salient structure with a more empirical parameterisation.

With this in mind, we write the monopole as 
\begin{align}
    P_0^{\mathrm{DS;s}}(k) = b^2_{\rm DS} b^2_\mathrm{g}\, {\cal K}_0(\beta_g)\,\big(P_0 + P_0^{NLC}\big) + P_{\rm stoch}, 
\end{align}
where 
\begin{align}
    P_0(k)=
    P_{\rm smear}(k)\,W^2(k,\alpha R),
\end{align}

\begin{align}
    P_0^{NLC}(k)=A_0\,
    \big(1-W(k,\alpha R)\big)\ \frac{P_{{\rm no wiggle}}(k)}{k^{n_0}},
\end{align}
\begin{align}
    P_{\rm stoch}(k) =C \,W^2(k,\alpha R).
\end{align}
Here, $b_{\rm DS}$ is the density-split bias factor discussed in the previous section; It is important for understanding the information content of DS clustering to be discussed in the next section. The factor $b_g$ is the Global linear bias of the full galaxy sample.  
${\cal K}_0 (\beta_g)= (1 + 2\beta_g/3 + \beta_g^2/5)$ is the additional bias factor which appears for the redshift-space monopole \citep{Kaiser1987}.  
If the linear theory power spectrum is 
$P_{\rm Lin}(k) = P_{\rm wiggle}(k) + P_{\rm no wiggle}(k)$
then 
$P_{\rm smear}(k)=P_{\rm nowiggle}+ \exp(-k^2\Sigma^2/2)P_{\rm wiggle}$. 
We set $\Sigma=5h^{-1}$Mpc to match the BAO feature in the monopole. 
The $P_0^{NLC}$ term is included to provide a correction to smeared linear theory; we allow it to have a free amplitude $A_0$ and large-$k$ slope (determined by $n_0$).  We use $W(k,\alpha R) = \exp(-(\alpha kR)^2/2)$, where $R$ is the scale on which the density splits were defined, and we allow $\alpha\ne 1$ when fitting.  In principle, $\alpha\ne 1$ arises from an average over Islands having a range of shapes, sizes and density profiles. (Indeed, a close look at Fig.~\ref{fig:2d_number_density} shows that the actual Island shapes will be more like doughnuts or torii of different sizes.)
The $P_{\rm stoch}$ term acts as a stochastic noise term at the largest scales (small $k$-values), which is likely related to the number density of the Islands.  

Note that the monopole also receives a shot-noise contribution, $P_{\rm SN}$, which is fixed by the number density of tracers in each DS bin. This only contributes to the smallest scales (the highest $k$-values). We always first subtract shot noise from the measurement before fitting. However, we retain the shot noise contribution in the covariance matrix used for the Fisher analysis as described in Section \ref{sec:fisher_forecast}.

For similar reasons, we write the quadrupole as 
\begin{align}
    P_2^{\mathrm{DS;s}}(k)= b^2_{\rm DS} b^2_\mathrm{g}\,{\cal K}_2(\beta_g)\, \big(P_2 + P_2^{NLC}\big),
\end{align}
with 
\begin{align}
    P_2(k)=
    P_{\rm smear}(k)\ W^2(k,\alpha R),
\end{align}
\begin{align}
    P_2^{NLC}(k)=A_2\,
    \big(1-W(k,\alpha R)\big)\ \frac{P_{{\rm no wiggle}}(k)}{k^{n_2}},
\end{align}
where ${\cal K}_2(\beta_g) = 4\,(\beta_g/3 + \beta^2_g/7)$.  We find no need for a $P_\text{stoch}$-like term for the quadrupole. 

\begin{figure}
\hspace{-0.1cm}
\centering
\includegraphics[width=0.48\textwidth]{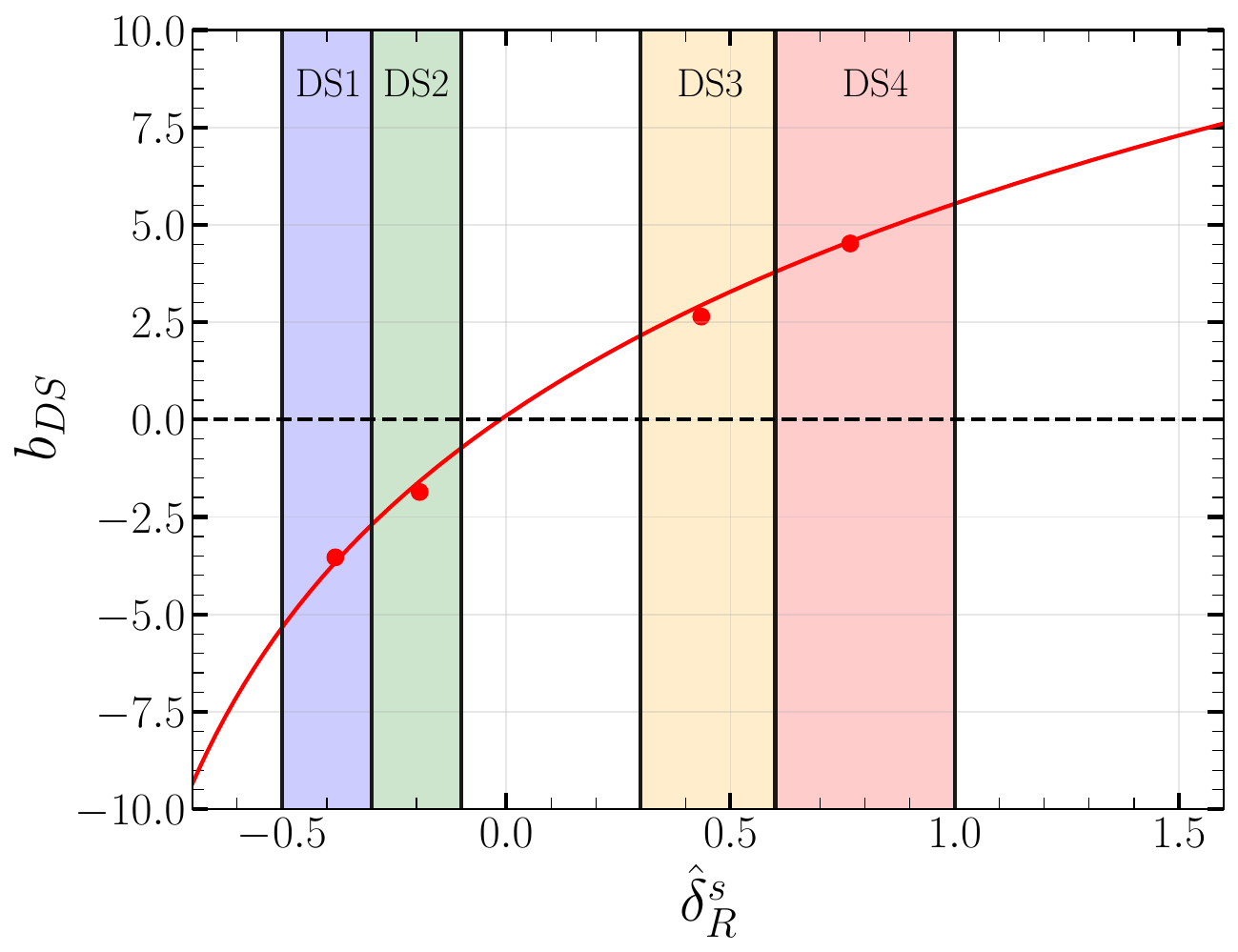}
\caption{The linear density-split bias $b_{\rm DS}$ in redshift space for the tracer-centred density PDF as a function of density. The PDF is from Fig.~\ref{fig:PDF_real_redshift_space}. $\delta_R^s$ represents the smoothed density with a Gaussian window of $R=5~h^{-1}$Mpc. The red data points are measurements from the approximate lognormal simulations at $z=2$. The vertical shaded regions of different colours represent the range of $\delta_R^s$ over which the averages are taken, corresponding to the four density bins (DS1-4) shown in Fig.~\ref{fig:PDF_real_redshift_space}. The red solid line is the model prediction.}
\label{fig:bias_prediction}
\end{figure}
Following the same spirit, one can also write down the model expression for the hexadecapole. Given that the cosmological information contained in the hexadecapole is relatively minor, we opt to neglect this component in this paper.

\begin{figure*}
\centering
\includegraphics[width=0.84\textwidth]{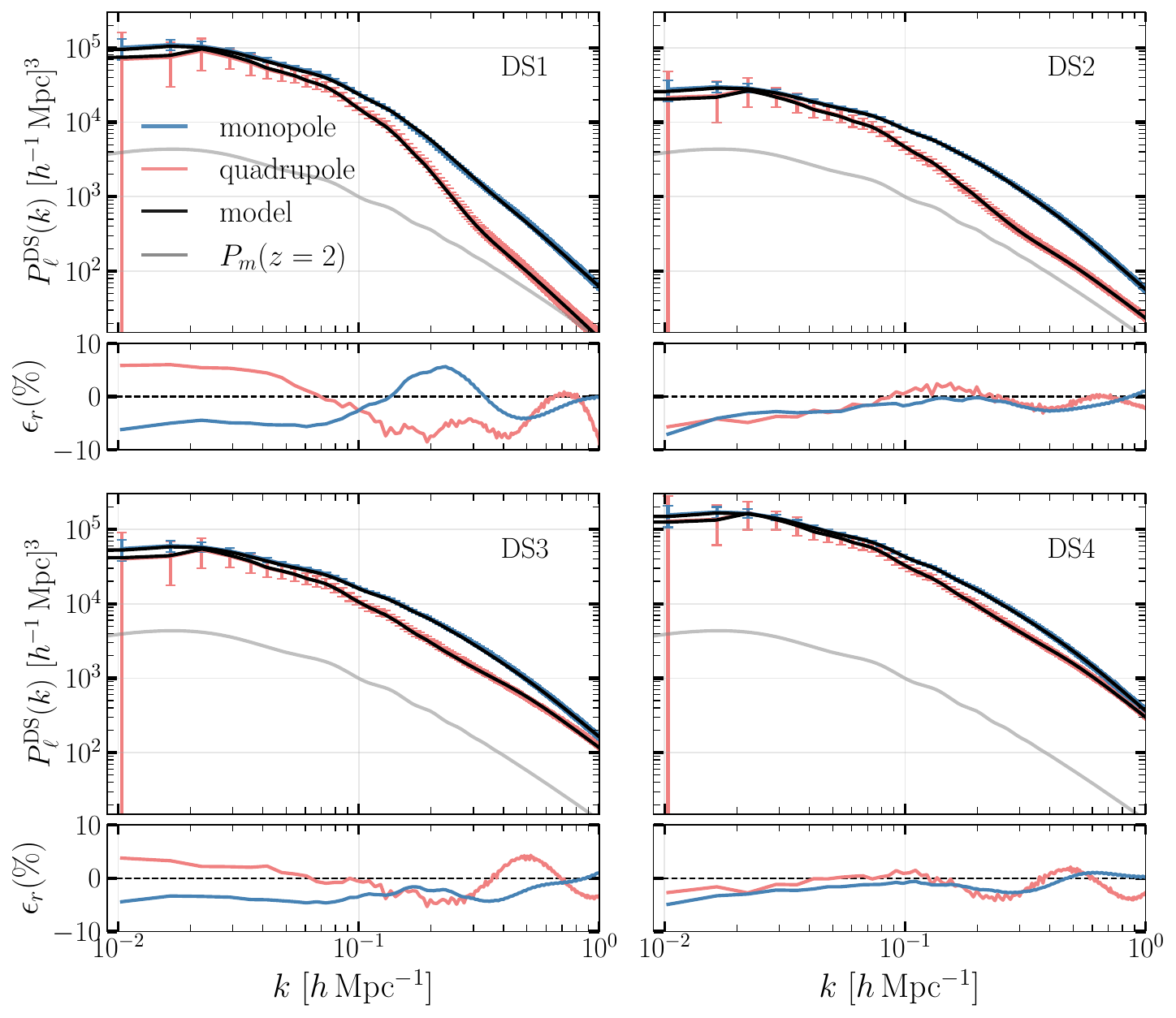}
\caption{Comparing the redshift-space power spectra measured from the approximate lognormal simulations (data points with errors, blue for the monopole, red for the quadrupole) with model predictions (solid black lines). The errorbars represent the 1$\sigma$ uncertainties from a sample of the number density of 6$\times 10^{-2} (h^{-1}\mathrm{Mpc})^{-3}$ in a volume of (1000~$h^{-1}$Mpc)$^3$. DS1-4 represent power spectra in the four DS bins shown in the corresponding colour regions of the tracer-centred PDF in Fig.~\ref{fig:PDF_real_redshift_space} at $z=2$ with a Gaussian smoothing scale of 5$~h^{-1}$Mpc. The grey solid line is the real-space matter power spectrum. It is the same for all panels as a reference for comparison. The ranges of the $x$ and the $y$ axes plotted are identical for the four panels. We see that the model matches closely to the measurement up to scales of $k\sim 1 h~\rm{Mpc^{-1}}$. The panels at the bottom show the relative error between the model and the measurement, $\epsilon_r = 100(P^{\rm model}_\ell - P^{\rm meas}_\ell)/P^{\rm model}_\ell$.}
\label{fig:quad_mono_Pk}
\end{figure*}
\begin{figure*}
\centering
\includegraphics[width=0.425\textwidth]{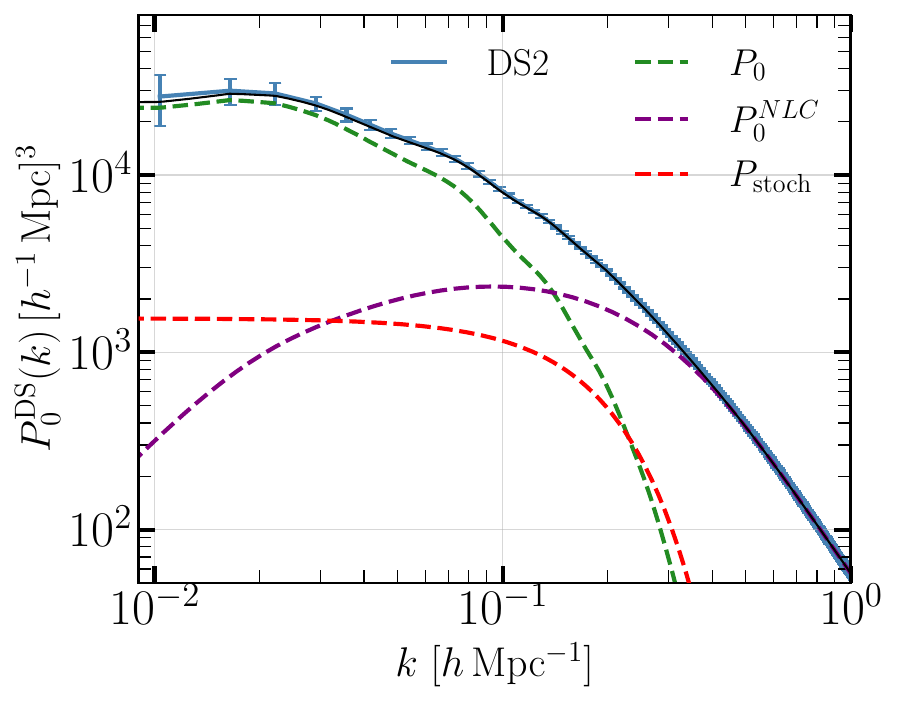}%
\hspace{-0.1cm}
\includegraphics[width=0.425\textwidth]{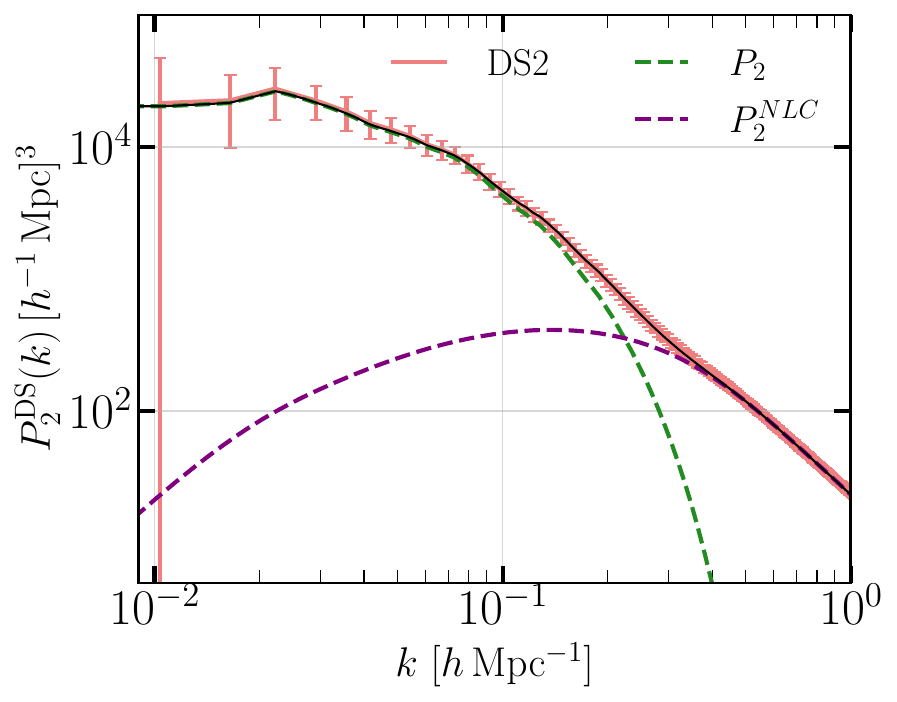}%
 \caption{An example of the contribution from the individual terms that constitute the monopole (left) and quadrupole (right) for DS2, the same as the top-right panel of Fig.~\ref{fig:quad_mono_Pk}. The sum of the contributing terms is shown in the black solid lines. Contributions of different terms are described in the main text and labelled in the legends.}
\label{fig:quad_mono_1_2_Island_term}
\end{figure*}
 
\begin{figure*}
\centering
\includegraphics[width=0.92\textwidth]{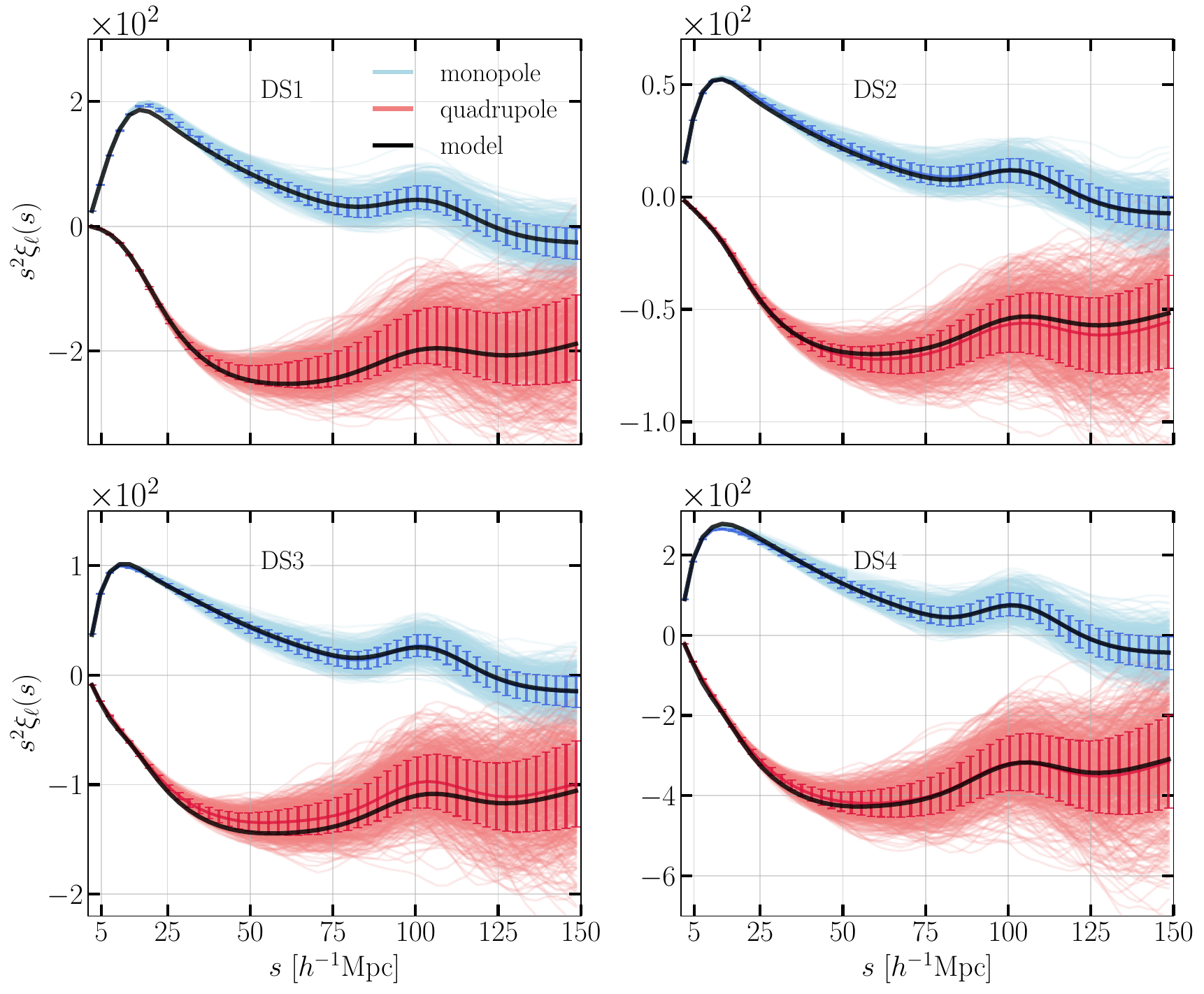}
\caption{The coloured data points with error bars show the measured monopole and quadrupole correlation functions for four different density bins. The blue and red curves are the 2000 {\sc mockfactory} realisations of the monopole and quadrupole correlation functions used to generate the covariance. The errorbars represent the 1$\sigma$ uncertainties from a sample of the number density of 6$\times 10^{-2} (h^{-1}\mathrm{Mpc})^{-3}$ in a volume of (1000~$h^{-1}$Mpc)$^3$.
The density bins correspond to the same 4 bins for which the power spectra are measured. In black lines, we show the model monopole and quadrupole correlation functions, which are the Fourier transforms of the model power spectra. }
\label{fig:monopole_quadrupole_correlation_function}
\end{figure*}
\begin{figure*}
\centering
\includegraphics[width=0.90\textwidth]{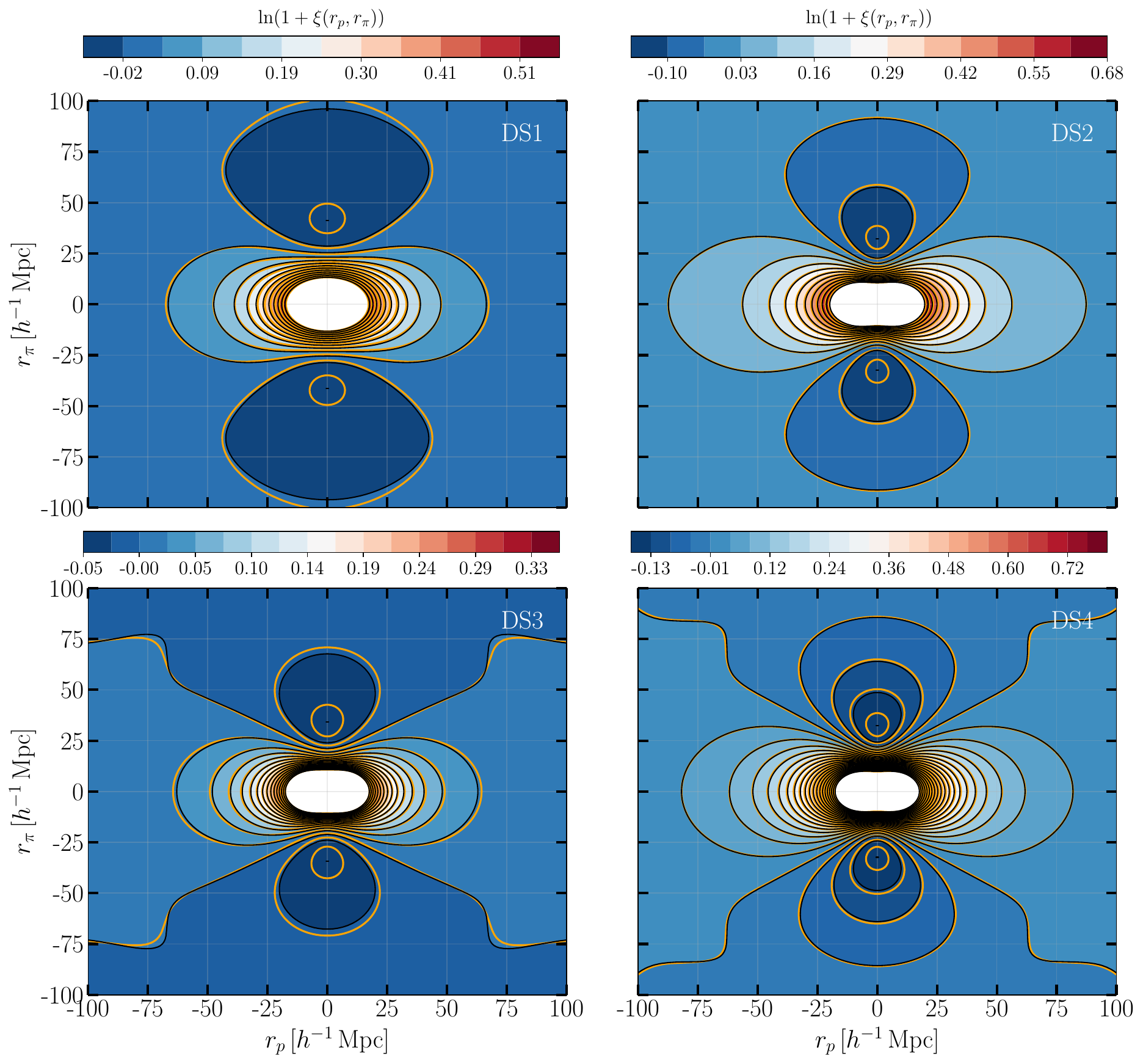}
\caption{Comparing 2D correlation functions in redshift space for the four different density bins (DS 1-4) between the approximate lognormal simulations (black) and model predictions (orange). The zoomed-in version of this figure for smaller scales is provided in Fig.~\ref{fig:2d_correlation_small_scale}. The measured 2D correlation functions are computed using the measured monopole and quadrupole power spectra. So, there is no hexadecapole information. The density bins correspond to the same density bins for which the monopole and quadrupole power spectra are measured.}
\label{fig:2d_correlation_large_scale}
\end{figure*}
Fig.~\ref{fig:bias_prediction} shows the analytic prediction for the DS bias with respect to $\delta_m$ using Eq.~\ref{DS_bias} along with the measured values from simulations. We see that the predicted DS biases match the simulation results quite well.
Fig.~\ref{fig:quad_mono_Pk} shows that this model is able to provide a good fit to measurements in simulations at the low-$k$ ($k \rightarrow 0.01~h\,\rm Mpc^{-1}$) limit for all DS bins. For a given DS bin, there are six free/nuisance parameters in the model, $A_0,\ A_2, \ n_0, \ n_2,\ C,$ and $\alpha$, which we determine as follows. First, we note that the variance of the overdensity field made from the tracers in each DS bin should satisfy 
\begin{align}
\label{constraint_eqn}
    (\sigma^{\rm DS}_{R})^2=\int \frac{d^3{\bf k}}{(2\pi)^3} \; P_0^{\mathrm{DS;s}}(k) W^2(k,R).
\end{align}

In practice, we use this to set $C$ for each DS bin. In general, the remaining five parameters would be obtained from a Markov Chain Monte Carlo (MCMC) fit of the model to the measured monopole and quadrupole power spectra. Here, however, as a demonstration of the formalism, we choose their values by visual inspection. We begin at the smallest scales (highest $k$), where the monopole and quadrupole are dominated by $P_0^{\rm NLC}$ and $P_2^{\rm NLC}$, respectively, as shown in Fig.~\ref{fig:quad_mono_1_2_Island_term}. On these scales, both multipoles behave like power laws, and matching their amplitudes and tilts gives $A_0, A_2, n_0$, and $n_2$. We then vary $\alpha$ to match the largest scales, with $C$ determined at each step by Eq.~\ref{constraint_eqn}, until the model gives a good fit to the measurements.

In summary, our model has five free parameters for each DS bin. The parameter $\alpha$ accounts for the variation in Island size across the different DS bins. The $A_\ell$ and $n_\ell$ parameters account for the amplitude and tilt of the non-linear power spectra.  They will be varied simultaneously with cosmological parameters during the fit with observational data or simulations. Marginalising over them will penalise the cosmological information we can obtain. We will now explore the performance of this model and compare the cosmological information we can obtain from the DS clustering with the conventional two-point statistics. 

\subsection{Testing the model with simulations}
\label{sec: model_simulation_test}
We test the performance of the model using mock galaxy catalogues described in Section~\ref{sec:simulation}. The measured redshift-space power spectra from galaxies in the four density bins are compared with the best-fit models in Fig.~\ref{fig:quad_mono_Pk}. We see that the amplitude of the monopole at low-$k$ ($k\sim 0.01~h\,\rm Mpc^{-1}$) is the highest for DS4, followed by DS1, DS3 and DS2. Before commenting on this ordering, we note that Fig.~\ref{fig:quad_mono_1_2_Island_term} shows how the various terms in our model add up to the total measured power spectra in one of the DS bins (DS2).  This shows that, on the largest scales, the power is mainly determined by the first term, $P_\ell$, whose amplitude (relative to other DS bins) is set by $b_\mathrm{DS}$.  Due to the linear responses of the $P_\ell$ term to the RSD and bias factors, which contain cosmological information, we anticipate it to be a relatively clean regime for exploiting cosmological constraints.
At high-$k$ ($k\sim 1~h\,\rm Mpc^{-1}$), the clustering is dominated by the $P_\ell^\mathrm{NLC}$ term. Until this term is well understood from first principles, we will have to marginalise over the nuisance parameters we use to describe it ($A_\ell, n_\ell, \alpha$ and the choice of smoothing window).  This will lead to a degradation of the cosmological constraints we derive.

The solid black lines in Fig.~\ref{fig:quad_mono_Pk} are the model predictions. The values of the nuisance parameters along with $C$ obtained in this fit are given in Fig.~\ref{fig:nuisance_parameters}. Comparing with the measurements from simulations, we can see that the model works well in capturing the scale dependence of both monopole and quadrupole over the plotted range. The accuracy of the model is at the level of a few per cent, as shown in the bottom panels of the figure.
As mentioned before, there are no complications from the Finger-of-God effect in our simulations, which need to be dealt with separately in future work using more realistic simulation setups.

The corresponding correlation functions are shown in Fig.~\ref{fig:monopole_quadrupole_correlation_function} (1D) and Fig.~\ref{fig:2d_correlation_large_scale} (2D), and are obtained by Fourier transforming the model spectra and comparing with the directly measured DS correlation function in redshift space. The 2D correlation functions are constructed from the monopole and quadrupole, so they do not carry any information from the hexadecapole. A zoomed-in version of this figure for smaller scales is provided in Fig.~\ref{fig:2d_correlation_small_scale}.
We can see that the models and the measurements agree very well in configuration space across the scales shown. The small-scale ($<25 \ h^{-1}\,\rm Mpc$) fit of the model in configuration space is sensitive to the fit of the power spectra at scales $\sim 0.1 h\,\rm Mpc^{-1}$, where the various terms in our model need to add up non-trivially to give the correct amplitude of the power spectra. These results suggest that the modelling framework works well.  
\begin{figure*}
\centering
\begin{overpic}[width=0.457\textwidth]{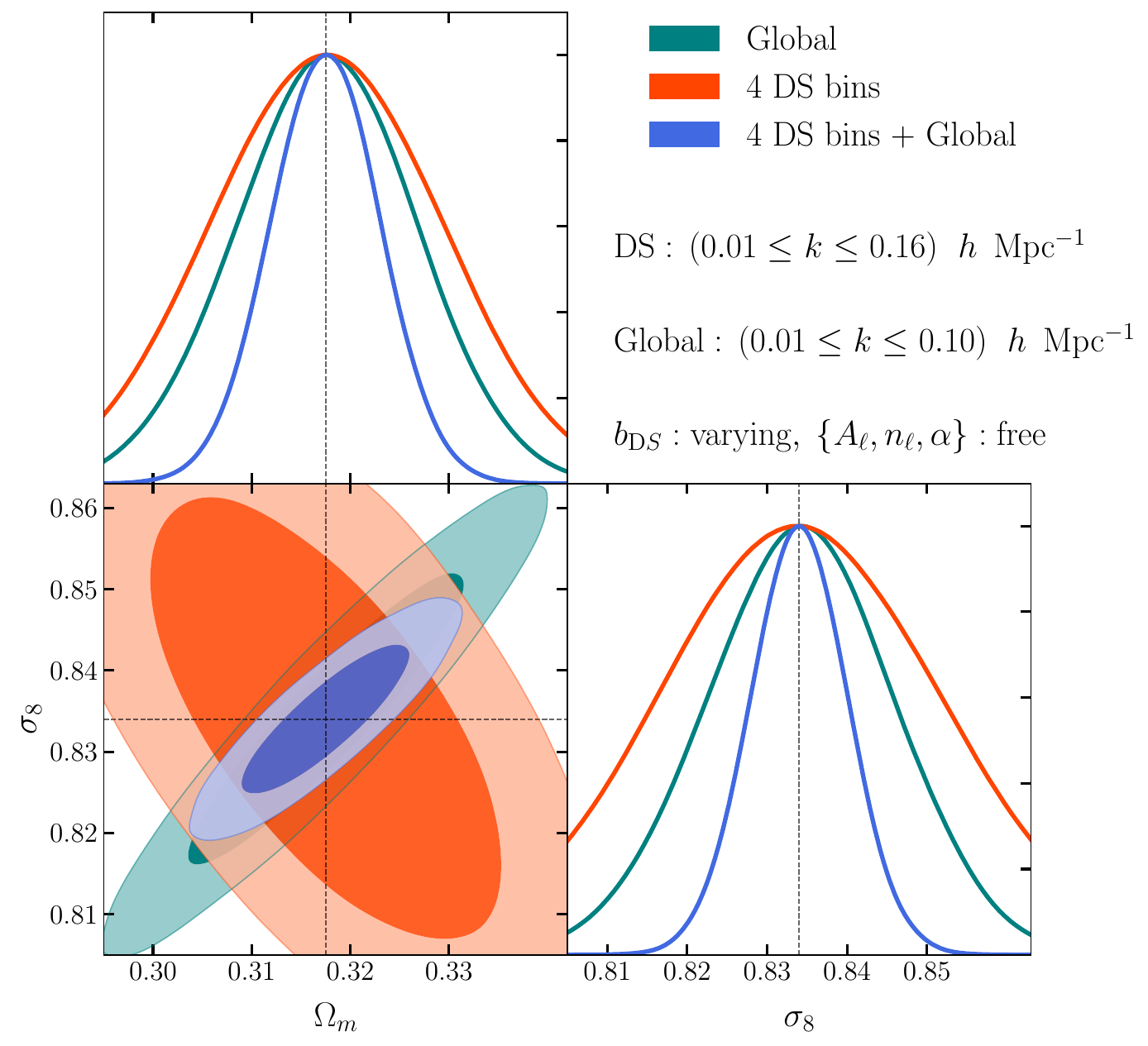}
  \put(13.5,82){\text{[A]}}
\end{overpic}%
\begin{overpic}[width=0.457\textwidth]{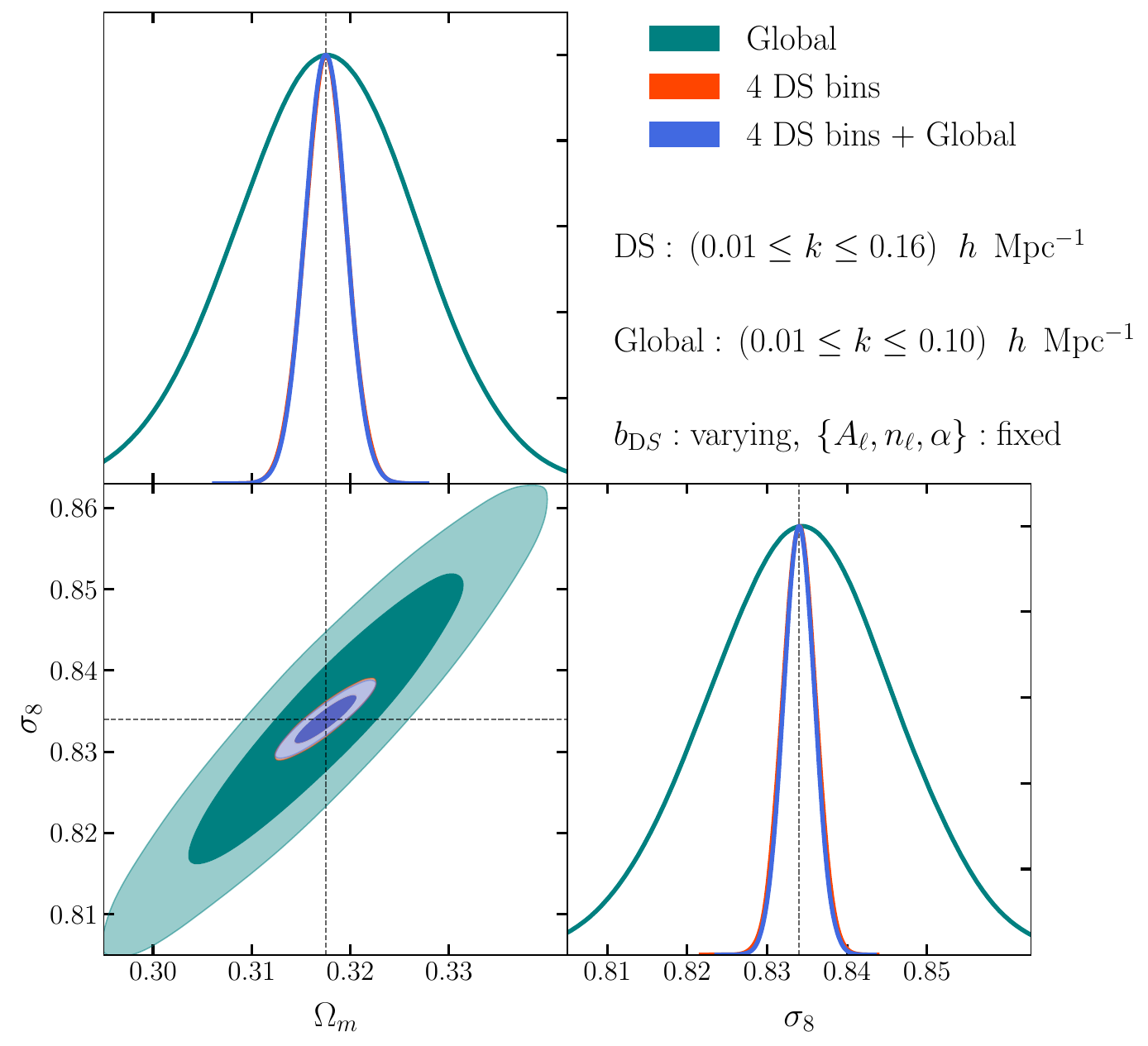}
  \put(13.5,82){\text{[B]}}
\end{overpic} \\%
\begin{overpic}[width=0.457\textwidth]{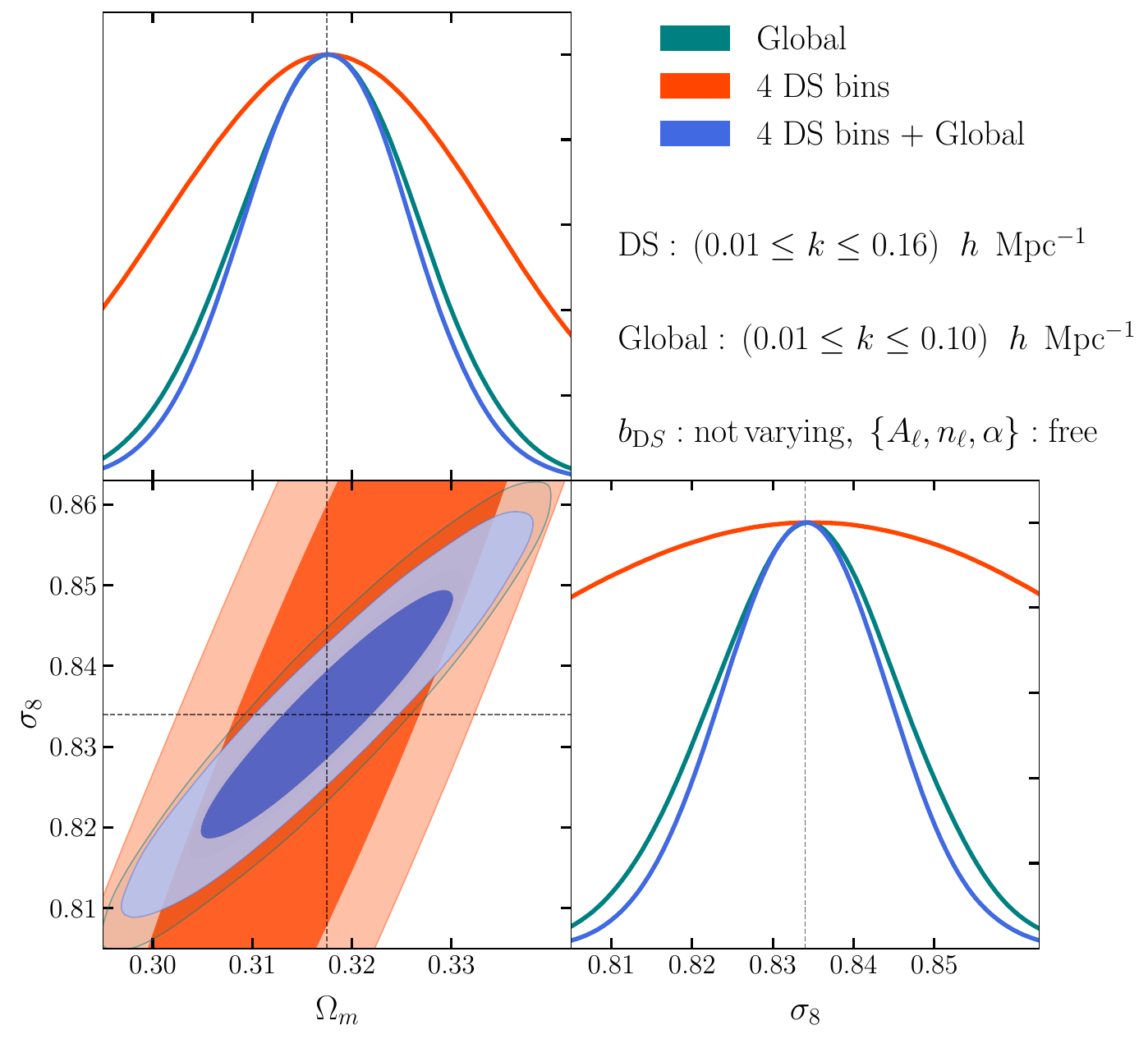}
  \put(13.5,82){\text{[C]}}
\end{overpic}%
\begin{overpic}[width=0.457\textwidth]{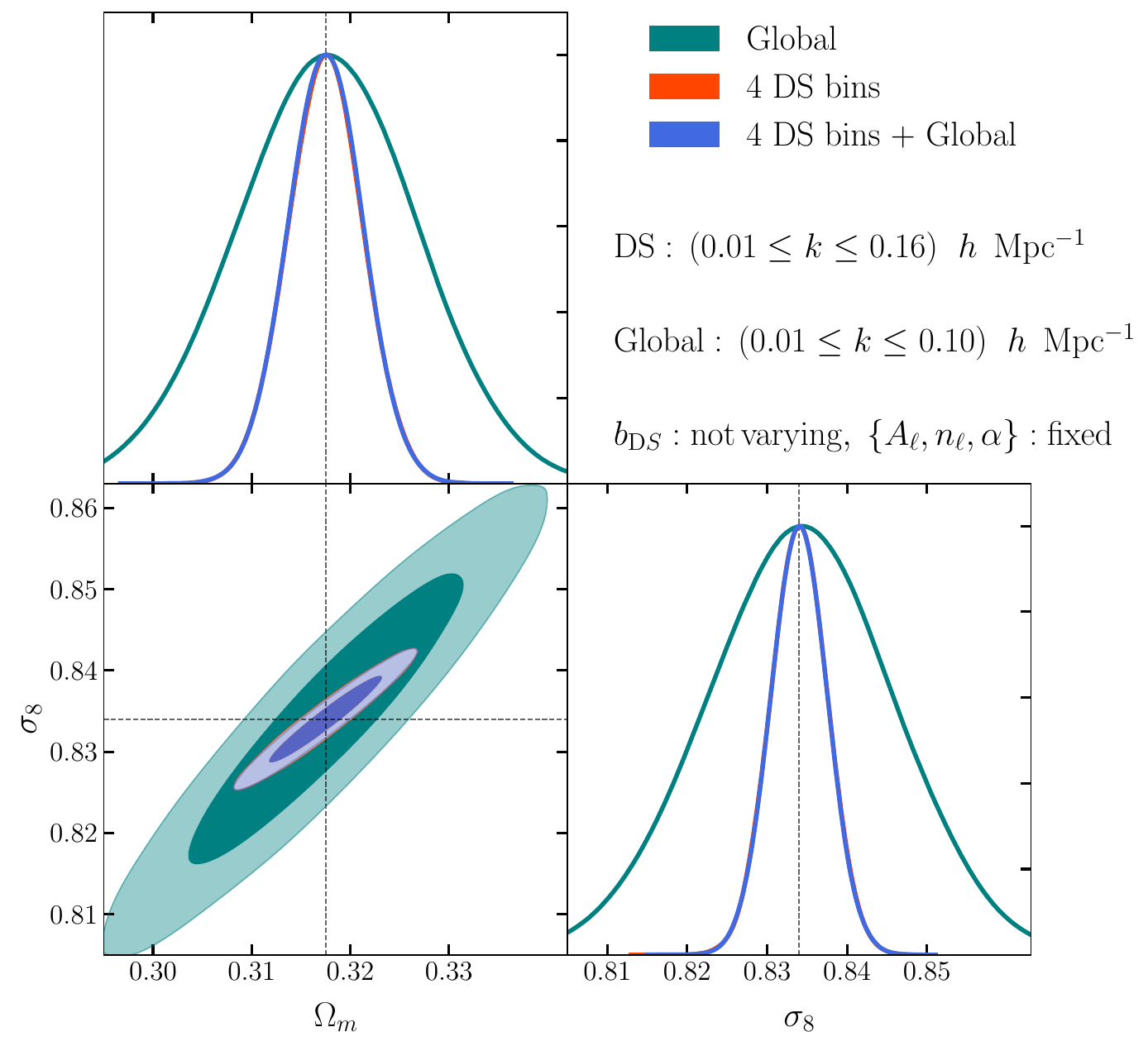}
  \put(13.5,82){\text{[D]}}
\end{overpic}
\caption{Comparing Fisher forecasts for constraining $\sigma_8-\Omega_m$ using the Global redshift-space monopole and quadrupole power spectra (green), the density-split clustering with four density bins (orange), and the combination of the two (blue). Ellipses show 68 and 95 per cent confidence levels from a sample of the number density of 6$\times 10^{-2} (h^{-1}\mathrm{Mpc})^{-3}$ in a volume of (1000~$h^{-1}$Mpc)$^3$. {\it Panel-A}: constraints when we marginalise over all 5 nuisance parameters in the model of DS clustering. The Global power spectra provide better constraints than DS clustering, but the direction of degeneracy is different; thus, their combination is better than the Global case alone. {\it Panel-B}: constraints with all 5 nuisance parameters fixed. In this case, DS clustering constrains the cosmological parameters better than the Global power spectrum alone. {\it Panel-C {\rm and} Panel-D}: These are repetitions of { Panel [A] and Panel [B]} cases, respectively, but here we have held the DS biases $b_{\rm DS}$ to their fiducial values, i.e. the DS biases do not change when we vary the cosmological parameters (note the broader range of values in the bottom left panel). We see that fixing the DS biases weakens the constraints from DS cases significantly. These two panels show that in DS clustering, a significant contribution to the total information content comes from the dependence of DS biases on cosmological parameters. The $k$-ranges used for the forecasts are shown in the figures. The details of the choice are explained in the main text. A companion figure with a common $k$-range for all cases is presented in Fig.~\ref{fig:Fisher_forecast_mockfactory_same_scale_cut}.}
\label{fig:Fisher_forecast_mockfactory_diff_scale_cut}
\end{figure*}
\section{Forecasts for constraints on cosmology}
\label{sec:fisher_forecast}
DS clustering has previously been demonstrated to be able to provide tighter constraints on cosmology than the Global 2PCF does \citep{Enrique2021, Paillas2023, Paillas2024MNRAS}, but the gap in modelling was filled with emulation or simplified assumptions. The understanding of {\it where the extra cosmological information may have come from in DS clustering} remains somewhat speculative. From the tracer-centred perspective, we have now demonstrated that our model can be accurate at the few-per-cent level, despite still bearing the linear assumptions. This makes it possible to see more transparently the connection between cosmology and the observables of DS clustering. We will now use it to explore the cosmological information in DS clustering and compare it with the Global two-point statistic. A main goal here is to provide a better understanding of the source of cosmological information in DS clustering.

We use the model described in Section \ref{sec: DS_cluster_model} to compute the Fisher information for DS clustering. We vary two cosmological parameters, $\Omega_m$ and $\sigma_8$, to test the response of the shape and amplitude of the power spectra. We also have 5 nuisance parameters per DS bin. Therefore, for the 4 DS-bins, we have a total of 22 parameters and a Fisher matrix of $22 \times 22$ dimensions,
\begin{align}
F_{ij} &= \sum_{\alpha,\beta} \frac{\partial d_\alpha}{\partial \theta_i} C^{-1}_{\alpha\beta} \frac{\partial d_\beta}{\partial \theta_j},
\end{align}
where $\theta_{i,j}=[\Omega_m, \sigma_8, A_0, A_2, n_0, n_2, \alpha]$. The data vector $d_{\alpha, \beta}$ is the monopole and quadrupole of the power spectra in linear $k$-bins, represented by the subscripts $\alpha$ and $\beta$. 

The covariance matrix is computed with 2000 realisations of the simulations using {\sc mockfactory},
\begin{align}
    C = \frac{1}{H_{\rm fac}}\left(\frac{1}{n_\mathrm{sim}-1}\right)\sum^{n_\mathrm{sim}}_{k=1}(\bm{d}_k-\bar{\bm{d}})(\bm{d}_k-\bar{\bm{d}})^{T},
    \label{eq:simple_covariance}
\end{align}
where $H_{\rm fac}=1-(n_{bin} + 1) / (n_{\rm sim} - 1)$ is the Hartlap correction factor \citep{Hartlap_2007}. The total length of the data vector $\bm{d}_k$ is given by $n_{\rm bin}$, and $n_{\rm sim}=2000$ is the number of realisations used for building the covariance matrix. We have verified that, for this data-vector length and number of realisations, the results of the Fisher information have converged.  
The Fisher forecast for the Global power spectra follows a similar procedure, with a linear biasing and RSD scheme, without any nuisance parameters. 
\begin{figure*}
 \centering
\begin{overpic}[width=0.33\textwidth]{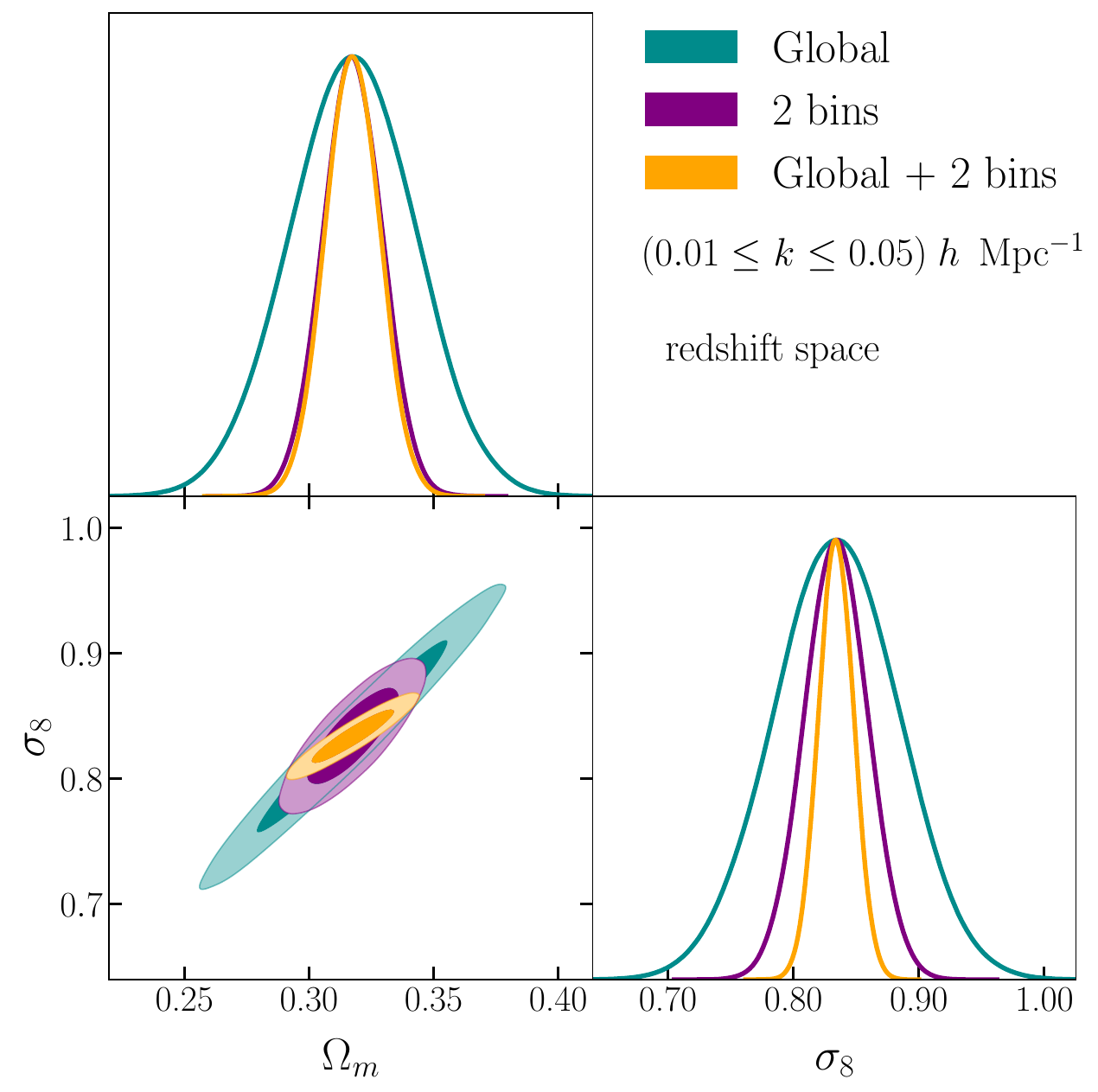}
  \put(13.5,92){\text{[A]}}
\end{overpic}%
\begin{overpic}[width=0.33\textwidth]{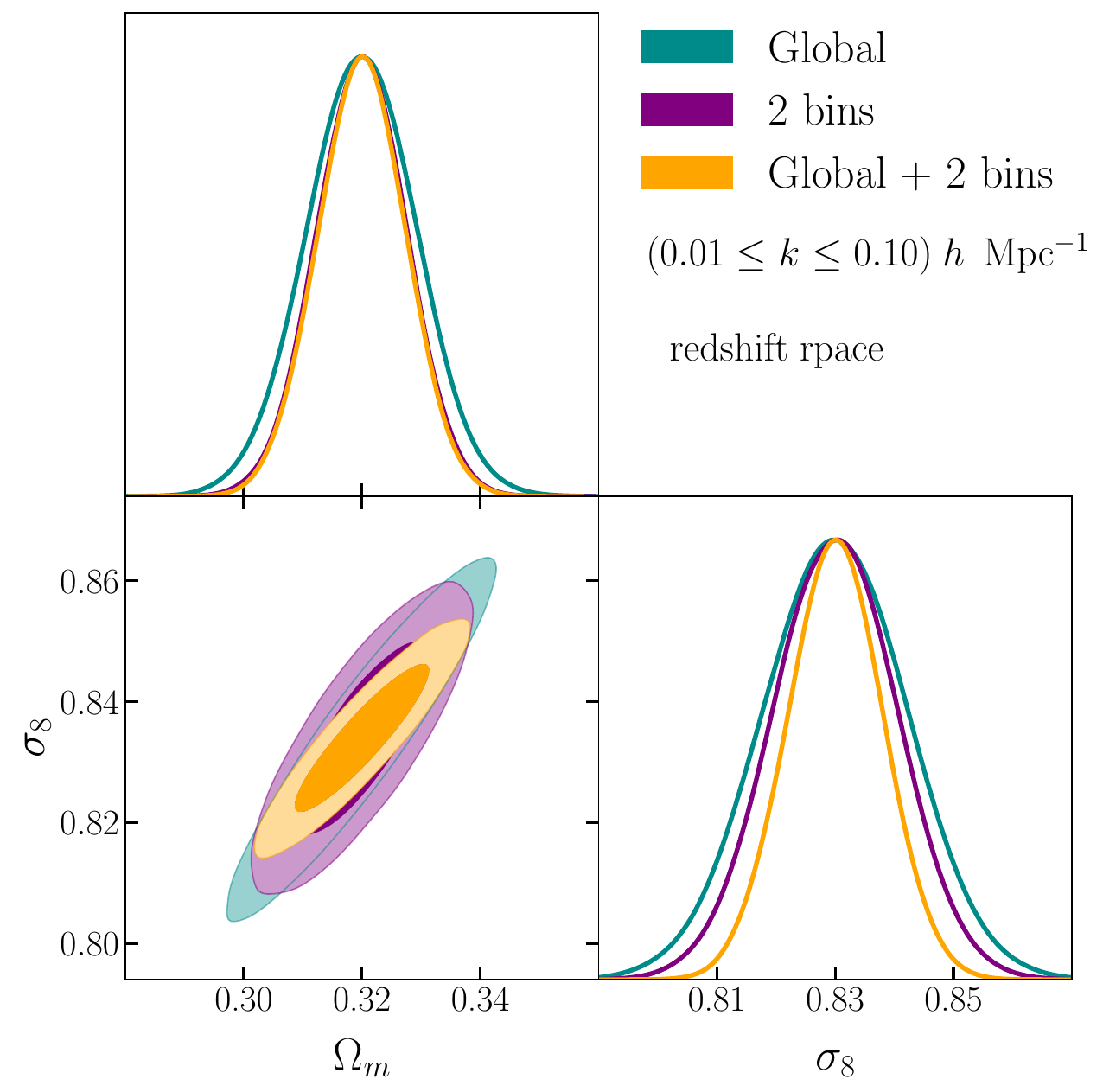}
  \put(13.5,92){\text{[B]}}
\end{overpic}%
\begin{overpic}[width=0.33\textwidth]{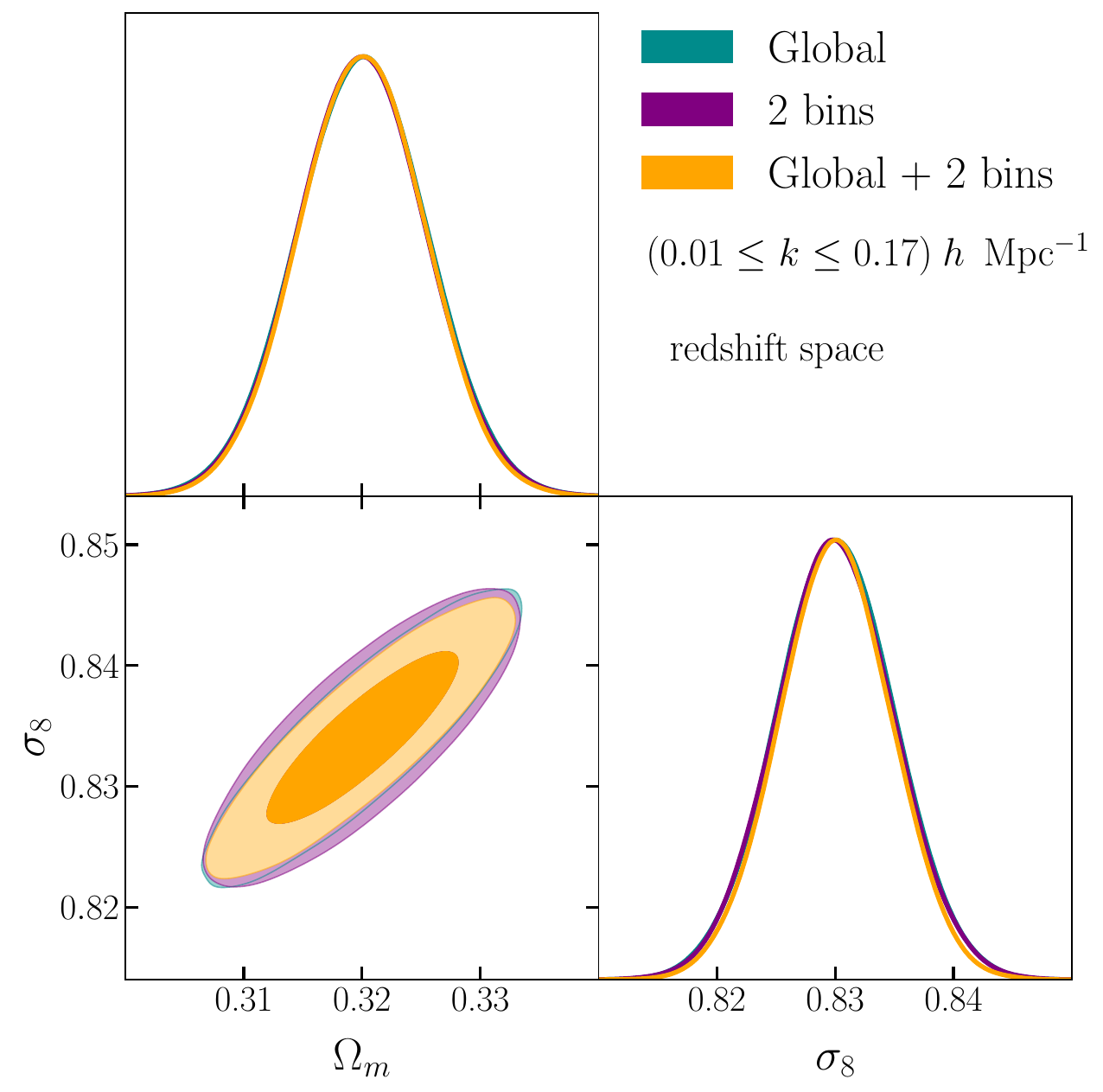}
  \put(13.5,92){\text{[C]}}
\end{overpic}
\caption{Comparing Fisher forecasts for the constraints of $\sigma_8-\Omega_m$ with the Global power spectra (Global, green), DS clustering (2 bins, purple), and the combination of the two (orange). From left to right are results with increasing $k_{\rm max}$, the maximal Fourier $k$ used in the forecasts. These are conducted with an almost perfectly Gaussian setting, with mock catalogues in redshift space at $z=10$ and a Gaussian filter with a smoothing scale of 10$~h^{-1}$Mpc. When $k_{\rm max}$ is small (Panel-A), the DS case has better constraining power than the Global case because the amplitudes of DS clustering contain contributions from perturbation modes larger than $k_{\rm max}$. As $k_{\rm max}$ increases (Panels B\&C), more high-$k$ modes are included in the Global case. This makes DS clustering and the Global case similar.} 
\label{fig: Fisher_information_Gaussian_pdf_z10}
\end{figure*}
We compare the constraints on $\Omega_m-\sigma_8$ for three cases: the Global power spectrum alone, DS clustering alone, and the combination of the two, shown in Fig.~\ref{fig:Fisher_forecast_mockfactory_diff_scale_cut}. We also present two extreme cases: with nuisance parameters marginalised over (panels A\&C), representing the conservative situation when the empirical model is adopted, and with nuisance parameters fixed (panels B\&D), which represent the case when we have a perfectly accurate model. Note that the Fisher analysis is done for fixed density bins, i.e. we keep the same values of density bins when we vary the cosmological parameters to calculate the derivatives. Also, the Global power spectra remain fixed when we measure the DS clustering. We also assume we have perfect knowledge of the linear bias. We present a more general case in Appendix \ref{app:free_linear_bias}.

Firstly, there is a strong positive correlation between $\Omega_m$ and $\sigma_8$ in the Global case. This is a consequence of our idealised \textsc{mockfactory} setup, where we have fixed the linear bias, $b_g=1$. For our choice of large- to moderate-scale $k$-modes, fixing the linear bias allows us to constrain the amplitude of the power spectrum directly through the $\Omega_m$ and $\sigma_8$ parameters. Increasing $\Omega_m$ decreases the amplitude of the power spectrum on large scales, which is compensated by a larger value of $\sigma_8$ to maintain the power spectrum at the same level. This results in a positive correlation in the degeneracy direction. Consistently, we checked that the degeneracy direction reverses when smaller scales ($k\gtrsim$ 0.4$\,h\,{\rm Mpc}^{-1}$) are included in the Global case. At smaller scales, the shift of the turnover to higher $k$ adds power at fixed $\sigma_8$. Thus, increasing $\Omega_m$ now raises the power and a lower $\sigma_8$ is required to compensate for that. Eventually, the large $k$-modes become the more dominant effect, and the degeneracy direction flips.

Note that we choose a more conservative $k$-cut for the Global power spectra than for the DS case. This is motivated by the fact that the perturbative approach of modelling the Global power spectrum breaks down at moderate k-values, while with the same requirement of accuracy, modelling of DS clustering may be able to reach smaller scales \citep{Enrique2021}. This is similar to the case presented in \citep{Repp2022}, where the maximum $k$-modes accessible to linear theory are shown to be higher in the indicator power spectra than in the Global one. We show the case where we use the exact same range of $k$ between the two cases in Fig.~\ref{fig:Fisher_forecast_mockfactory_same_scale_cut}.
Next, we see that when all the nuisance parameters are marginalised over, the constraining power of DS clustering alone is substantially degraded compared to the Global case. However, the direction of degeneracy for $\Omega_m-\sigma_8$ is very different between these two cases.
This is because of the cosmological dependence of $b_{\rm DS}$ through the tracer-centred PDF. We first consider Panel A in Fig.~\ref{fig:Fisher_forecast_mockfactory_diff_scale_cut}. Our bins are defined at fixed density thresholds, and their biases show how extreme those bins are compared to the variance of the smoothed field. We can approximate, $b_{\rm DS}\simeq (\delta_{\rm DS}-\bar\xi(R))/\sigma^2(R)$. On large scales, this means that increasing the value of $\sigma_8$ will increase $\sigma^2(R)$. $b_{\rm DS}^2$ falls faster than $P(k)$ rises, and so the amplitude of DS clustering decreases. Decreasing $\Omega_m$ at fixed $\sigma_8$ increases $P(k)$ on large scales, but also increases $\sigma^2(R)$, which lowers $b^2_{\rm DS}$; on the largest scales, the former dominates. Therefore, this leads to a preference for lowering $\Omega_m$ to restore the value of the power spectrum. This explains the anti-correlation in the degeneracy direction in the $\sigma_8-\Omega_m$ plane. Since we are marginalising over the nuisance parameters, there is almost no information coming from small scales.
So, their combination provides tighter constraints on cosmological parameters than does the Global case alone. Note that we used the joint covariance for the combined analysis.  

When the nuisance parameters are fixed to their fiducial values (Panel B), the DS clustering becomes significantly more constraining than the Global power spectra. 
This suggests that the freedom allowed in the $P_\ell^{NLC}$ term degrades the constraints on cosmology. Fixing the nuisance parameters restores the information on intermediate scales. Increasing $\Omega_m$ lowers $\sigma^2(R)$, which raises $b_{\rm DS}^2$, by the same fractional amount on all scales, whereas its effect on $P(k)$ weakens. $P(k)$ therefore barely changes while $b_{\rm DS}^2$ still increases, so the amplitude of DS clustering now increases with $\Omega_m$. Since $b_{\rm DS}$ now decreases with $\sigma_8$, we need to increase $\sigma_8$ to compensate. This explains the reversal of the degeneracy direction. We explicitly checked that, if we restrict to even larger scales $k<0.04 \,h\,{\rm Mpc}^{-1} $, we do not see a reversal of the degeneracy direction even if we fix the nuisance parameters. This supports our reasoning above.
This comparison is intentionally optimistic, but it illustrates the great potential of DS clustering if the modelling of this term becomes more accurate, which will be studied in future work. 

We suspect that the extra information is coming from the DS bias entirely. We test this by fixing $b_{\rm DS}$. Note that $b_{\rm DS}$ is never an unknown free parameter of the model. When $b_{\rm DS}$ is mentioned as `varying', we mean that $b_{\rm DS} $ responds to changes in cosmology. $b_{\rm DS}$ in that case is obtained from the tracer-centred PDF. $b_{\rm DS}$ `not varying' means that we have artificially held it to its fiducial value to find out how much information DS clustering would carry if the DS biases do not respond to cosmology. We find that fixing $b_{\rm DS}$ leads to no extra information from DS analysis compared to the Global case. These are cases shown in panels C\&D, where we see that the constraints on cosmology degrade considerably from panels A\&B for the DS cases. However, in Fig.~\ref{fig:Fisher_forecast_mockfactory_diff_scale_cut}, this is difficult to visualise due to the different scale cuts. 
It is better shown in panels C\&D in Fig.~\ref{fig:Fisher_forecast_mockfactory_same_scale_cut}, where the k-ranges are the same between DS-clustering and the Global case. We see in Panel D that with the same scale cut, there is no extra gain of information in the DS case, even when the nuisance parameters are fixed.

In summary, we confirm that there is a significant gain of cosmological information with DS clustering when the model has no extra nuisance parameters compared to the Global power spectra, just as seen in \citet{Paillas2023}. When the 20 nuisance parameters are marginalised over, DS clustering becomes less constraining than the Global power spectrum. However, their combination can still tighten the constraints more than using the Global power spectrum alone due to degeneracy breaking, when DS clustering is analysed to smaller scales than the Global power spectra.  
\subsection{Where does the information come from?}
The difference in the directions of the ellipses between DS-cluster and Global power spectra presented in panel-A of Fig.~\ref{fig:Fisher_forecast_mockfactory_diff_scale_cut} suggests that there must be extra cosmological dependence in DS clustering. Given that the linear bias of tracers $b_g$ and linear RSD are common to the two cases, the extra information must come from $b_{\rm DS}$. In other words, the amplitudes of DS clustering carry cosmological information via $b_{\rm DS}$. Indeed, as we have shown in Section~\ref{sec:PDF}, $b_{\rm DS}$ contains information from the (non-Gaussian) density PDF, which incorporates perturbation modes at high-$k$ through the variance $\sigma^2(R)$ and $\bar\xi(R)$ below the smoothing scale of $R$. Crucially, such a dependence is different from that of the matter spectrum. These are the main reasons for the ellipses of $\Omega_m-\sigma_8$ to rotate relative to the Global case. 

To further illustrate this point, we make a test case of Fisher forecasts for a nearly perfect Gaussian field at $z=10$, shown in Fig.~\ref{fig: Fisher_information_Gaussian_pdf_z10}. We compare the Global case with a 2-bin-DS clustering case, where we include the cross-power spectra in our analysis. The inclusion of the cross-bin power spectra ensures that the total number of pairs measured in DS clustering is identical to the Global case. So, compared to the Global power spectrum, DS clustering amounts to re-binning of the galaxy pairs according to their environmental density. Here, we use the mean of 2000 different realisations as our data vector. The derivatives of the data vector with respect to the cosmological parameters are also calculated using simulations. For a Gaussian field, we expect that the DS clustering and the Global power spectra should offer the same cosmological constraints if the Fourier modes involved are the same. However, in DS clustering, since we smooth the field with a Gaussian kernel of a finite size $R$, some perturbation modes at  $k\sim 1/R$ still contribute to $\sigma^2(R)$ and $\bar\xi(R)$, and hence also to $b_{\rm DS}$. So, when we have a relatively small $k_{\rm max}$ (left-hand panels of the figure, panel A), the Global power spectrum loses all the modes greater than $k_{\rm max}$, while DS clustering retains some high-$k$ modes through $b_{\rm DS}$. So, DS clustering offers better constraints despite having the same $k$-range as in the Global case. When we have a large $k_{\rm max}$, however, the high-$k$ modes are also included in the Global power spectra, and so the constraints on $\Omega_m-\sigma_8$ are nearly identical to DS clustering (right-hand panels of the figure, panel C). These results are consistent with the tests conducted by \citet{Paillas2023} in configuration space. 

So, in DS clustering, the cosmological dependence enters in two places. First, it enters through the usual matter power spectrum and RSD factors. Second, it enters through the predicted tracer-centred PDF and therefore through the DS bias $b_{\rm DS}$. The latter is important because the PDF depends on integrals over the smoothed power spectrum, so information from modes near the smoothing scale is retained even when the final clustering data vector is cut at a larger scale. Additionally, a non-Gaussian PDF also carries information about all higher moments, which is absent for the Global power spectra.  
\section{Conclusions and discussions}
\label{sec:conclusion}
Density-split (DS) clustering is a powerful tool to extract cosmological information beyond the two-point statistics. However, connecting the modelling of the DS power spectra to observational data has been challenging. In this work, we present a formalism to model the DS clustering directly in redshift space. Although the model has been validated using an idealised simulation setup, it demonstrates key concepts required to achieve this goal. First among these is the conservation of DS voxels, achieved by evaluating the number density only at the location of the galaxies. The outcome is a tracer-centred density PDF, rather than the volume-weighted PDF from the normal counts-in-cells (CIC). We provide a step-by-step procedure for mapping the traditional CIC-smoothed density field to the tracer-centred density field in redshift space. This is crucial for calculating the linear bias associated with environmental densities, $b_{\rm DS}$. In essence, we can now predict the amplitude of DS clustering in redshift space for a given cosmology and smoothing window. We then use this to construct a generalised halo model for the DS power spectra in redshift space, consisting of the one-Island and two-Island terms.
The two-Island terms on large scales are linearly mapped to the matter power spectrum and proportional to $b_{\rm DS}$. We parameterise the one-Island term with nuisance parameters.  

The current model is empirical, but it demonstrates that the modelling of large-scale clustering in DS statistics can be achieved in the same way as the normal two-point statistics. The idea of number weighting makes the environmental density act as a tag (or weight), assigned to each galaxy. So, DS clustering is, indeed, a marked statistic \citep{Sheth2005MNRAS, White2016JCAP}.  

We use the model to make Fisher forecasts for cosmological constraints and compare them with the case of Global power spectra. We confirm that DS clustering can provide tighter constraints than the normal two-point statistic, if the accuracy of the model is under control, as seen in \citep{Paillas2023}. We confirm that this is still the case when the field is Gaussian, which seems counterintuitive. We identify that the reason resides in the extra cosmological information contained in $b_{\rm DS}$: The DS bias depends on the tracer-centred PDF, which in turn depends on the matter power spectra up to very high-$k$ modes through the variance $\sigma^2(R)$ and $\bar{\xi}(R)$. So, despite a face-value high-$k$ cut-off in the clustering analysis, DS power spectra contain information from $k$-modes beyond the cut-off via $b_{\rm DS}$ encoded in the amplitude of DS clustering on large scales. When there is no scale cut-off in the clustering analysis, DS clustering and the Global power spectra include the same modes, and therefore the constraints on cosmology are the same.

The present model includes five nuisance parameters per density bin, which weakens the constraining power on cosmological parameters once marginalised over. Despite this loss, we show that combining the DS power spectra with the Global case can yield additional cosmological information, owing to the differing degeneracy directions of the two probes. 

\subsection{Connections to other beyond-two-point statistics}
\label{sec:indicator_field}
As mentioned before, binning through a tracer-centred density PDF turns DS clustering into a marked statistic. By associating each galaxy with an environmental density and splitting them into density bins using a threshold, we essentially assign weights of $[0, 1]$ to each galaxy. As long as we count galaxy pairs across different density bins, the total number of pairs matches the standard two-point correlation function. We could also use other galaxy properties, such as luminosity or host halo mass, as indicators for the split and assign them as weights. In this sense, DS clustering is a special case of mark correlations \citep{Beisbart2000ApJ, Sheth2005MNRAS}. Moreover, because the lognormal mapping in Eq.~\ref{DS_bias} is monotonic, our DS clustering is a special case of the `rank ordered' mark correlation analyses described in \cite{Skibba2013}.

DS clustering is closely related to indicator power spectra \citep{Repp2022}. We describe the indicator field in real space using volume-weighted splitting. If we follow our prescribed method for measuring power spectra but, instead of using tracer-centred measurements, use the volume-weighted PDF in real space, we can define an equivalent real-space indicator field with the same power spectrum. This field is analogous to the one described in \citep{Repp2022}, with additional smoothing using the same chosen window function for the definition of the density bins as follows:
\begin{align}
\label{indi_func}
    I=\bigg\{ {\genfrac{}{}{0pt}{}{(1+\hat\delta^s) \;\;\rm if \; \; \hat\delta^s_R \in DS} {0\;\;\;\;\,\rm otherwise,}}
\end{align}
where
\begin{align}
\label{del_indi_func}
    \delta_I =I/\langle I \rangle-1.
\end{align}
Thus, if the density bins are narrow enough, the correspondence is through volume weighting and tracer-centred density bins. If  $I_\mathrm{RS}$ denotes the indicator functions described in \citep{Repp2022}, and $I$ those in Eq. \ref{indi_func}, then 
\begin{align}
    I_{} &= (1+\delta^s)\cdot I_\mathrm{RS},\\
    \langle I_{} \rangle &\approx (1+\delta^s) \langle I_\mathrm{RS}\rangle,
\end{align}
for narrow bins, so that
\begin{align}
    \delta_{I_{}} = \frac{I_{}}{\langle I_{} \rangle} - 1 = \frac{I_\mathrm{RS}}{\langle I_\mathrm{RS} \rangle} -1 = \delta_{I_\mathrm{RS}}.
\end{align}
Thus, any difference in power spectra reflects either (1) a difference in smoothing strategy, (2) tracer-centring vs. volume-weighting, and/or (3) the impact of finite bin width.

We tested that, with the above treatment of the field, we recover the same density-split power spectra as defined in this paper. Thus, density-split clustering is essentially the same as the clustering of the weighted indicator field as defined in Eqs.~\ref{indi_func} \& \ref{del_indi_func}. 
\section*{Data availability}
{\sc mockfactory} \url{https://github.com/cosmodesi/mockfactory} is free software distributed under a BSD3 license. The simulations, codes, and results are available upon reasonable request to the authors.
\section*{Acknowledgments}
YC acknowledges the support of the UK Royal Society through a University Research Fellowship.
This research was partly supported by the Munich Institute for Astro-, Particle and BioPhysics (MIAPbP), which is funded by the Deutsche Forschungsgemeinschaft (DFG, German Research Foundation) under Germany´s Excellence Strategy – EXC-2094 – 390783311. IS acknowledges NASA ROSES grants 80NSSC24K1489 and 24-ADAP24-0074, and contract number 80NM0018F0610 via a JPL sub-award.  RKS is grateful to the Institute for Advanced Studies at the University of Bologna for its hospitality in Fall 2025, to IUCAA for hospitality in Winter 2026, and to the University of Genova and the University of Arizona for hospitality in Spring 2026.  
AG thanks Shadab Alam for useful discussions.

\bibliographystyle{mnras}
\bibliography{references}
\appendix
\section{Derivation of the tracer-centred PDF}
\label{Appendix_TC_PDF}
The starting point is to assume there is a continuous field $\delta_g({\bf x})$.  For example, one might make it via a lognormal transform of the continuous field $\delta_m({\bf x})$.  We would like to turn this into a discrete point process.  One way to do this is called Poisson sampling, in which the probability of having a galaxy at ${\bf x}$ is given by a Poisson distribution with mean proportional to $1+\delta_g({\bf x})$.  We will call this `local' sampling, as it does not depend on $\delta_g$ at any other position.  If this mean is sufficiently small, then each point in space receives a value of $N_g({\bf x})=1$ or 0, with probability $p_1=\bar{N}_g \exp(-\bar{N}_g)\to \bar{N}_g$ and $p_0=\exp(-\bar{N}_g)\to 1 - \bar{N}_g = 1-p_1$.  For example, if we wish to produce a point process with number density $\bar{n}_g$, then 
$$p_1 = \bar{N}_g = \bar{n}_g [1+\delta_g({\bf x})]\,d{\bf x}$$ 
for a small volume element $d{\bf x}$ around position ${\bf x}$.

Top-hat smoothing corresponds to integrating $d{\bf x}$ over a volume $V$ with each $d{\bf x}$ receiving the same weight, so it amounts to simply summing up the discrete counts within $V$.  The sum of independent Poisson variates, each having mean $\bar{N}_i$, is itself a Poisson variate, with mean given by 
$\sum_i \bar{N}_i$.  For our case, the mean value for a top-hat sphere of radius $R$ centred on position ${\bf x}$ becomes $\bar{n}_gV\,[1+\delta_{g,R}({\bf x})]$, where $\delta_{g,R}$ denotes the volume average of the underlying continuous field at this position: 
 $\delta_{g,R}({\bf x})=\int d^3{\bf y}\, W(|{ \bf x-y}|,R)\,\delta_g({\bf y})$ (note that $W$ has been normalised so that its volume integral is unity.)

Thus, the probability of finding $N$ galaxies within a top-hat filter of radius $R$ that is placed randomly in an underlying galaxy overdensity field, $\delta_g$, is given by
\begin{align}
    P(N)=\int d\delta_{g,R} \,P_{\rm Poisson}(N|\delta_{g,R})\,P(\delta_{g,R}).
\end{align}
If $\delta_g=b_g\,\delta_m$ then $P(\delta_{g,R}) = P(\delta_{m,R})/b_g$, 
when the fields are smoothed by $R$. To reduce clutter we will denote $\delta_{g,R}$ as $\delta_{R}$. The term $P_{\rm Poisson}(N|\delta_R)$ describes the Poisson sampling of the underlying smooth galaxy field. Thus, $P_{\rm Poisson}(N|\delta_R)$ is a Poisson distribution with mean $\lambda=\bar{N}(1+\delta_R)$, where $\bar{N}=\bar{n}_gV_R$. Here $\bar{n}_g$ is the chosen galaxy number density and $V_R$ is the volume within the smoothing scale, for a top-hat filter. 

Next, we want the conditional probability of finding $N$ galaxies in $V$, given that $V$ is not randomly placed, but is centred on a galaxy, i.e. we want to find $P(N|{\bf x_g})$. Thus,
\begin{align}
\label{P_N_given_xg}
    P(N|{\bf x_g})= \int d\delta_R \,P_{\rm Poisson}(N|\delta_R,{\bf x_g})\,P(\delta_R|{\bf x_g}).
\end{align}
Consider the second term, $P(\delta_R|{\bf x_g})$. Using Bayes' theorem, we can write,
\begin{align}
\label{P_deltaR_given_xg}
    P(\delta_R|{\bf x_g})\propto P({\bf x_g}|\delta_R)P(\delta_R).
\end{align}
In the Poisson sampling model, the probability of finding a galaxy at a position ${\bf x}_g$ is what we previously called $p_1\propto 1+\delta_g({\bf x_g})$. 

Now we want the conditional probability $P({\bf x_g}|\delta_R)$:  given that the smoothed density is $\delta_R$, what is the probability that there is a galaxy at the centre? Thus,
\begin{align}
    P({\bf x_g}|\delta_R)&\propto \langle(1+\delta_g)|\delta_R\rangle \nonumber\\
    &\propto 1+ \langle \delta_g|\delta_R\rangle.
\end{align}
Assuming a Gaussian  distribution for the conditional probability distribution, we get,
\begin{align}
    \langle\delta_g|\delta_R\rangle = \frac{\langle \delta_g\delta_R\rangle}{\langle \delta^2_R\rangle}\delta_R=\frac{b^2_g\,\bar{\xi}(R)}{b^2_g\,\sigma^2(R)}\delta_R = \frac{\bar{\xi}(R)}{\sigma^2(R)}\delta_R
\end{align}
Therefore, 
\begin{align}
\label{second_term_final}
   P(\delta_R|{\bf x_g})\propto\left(1+\frac{\bar{\xi}(R)}{\sigma^2(R)}\delta_R\right)  P(\delta_R),
\end{align}
so
\begin{align}
  P(\delta_R|{\bf x_g})=   \frac{\left(1+\frac{\bar{\xi}(R)}{\sigma^2(R)}\delta_R\right)  P(\delta_R)}{\int d\delta_R \left(1+\frac{\bar{\xi}(R)}{\sigma^2(R)}\delta_R\right)  P(\delta_R)}
\end{align}
Now, let us look at the Poisson term, i.e. $P_{\rm Poisson}(N|\delta_R,{\bf x_g})$. 
\begin{align}
    P_{\rm Poisson}(N|\delta_R,{\bf x_g})={\rm Poisson}((N-1)|\lambda).
\end{align}
This is because we are fixing one galaxy, so the rest of the galaxies are Poisson distributed. Therefore, in the limit of large $\lambda$, i.e. for large $\bar{n}_g$, we have 
\begin{align}
    {\rm Poisson}((N-1)|\lambda) \simeq \frac{1}{\sqrt{2\pi\lambda}}\exp\left(-\frac{(N-1-\lambda)^2
    }{2\lambda}\right).
\end{align}
Doing the change of variable, $N/\bar{N}-1=\hat\delta_R$, we get $N-1-\lambda=\bar{N}(\hat\delta_R-\delta_R-\frac{1}{\bar{N}})$. Therefore, in this limit we get,
\begin{align}
    P_{\rm Poisson}(N|\delta_R,{\bf x_g})&\equiv P_{\rm Poisson}(\hat\delta_R|\delta_R,{\bf x_g})=\nonumber \\
    &\sqrt{\frac{1}{2\pi(1+\delta_R)/\bar{N}}}\exp\left(-\frac{(\hat\delta_R-\delta_R-\frac{1}{\bar{N}})^2}{2(1+\delta_R)/\bar{N}}\right).
\end{align}
Now, finally, in the limit of large $\bar{N}$, i.e. for the large R limit, the variance shrinks, and the Gaussian distribution goes towards a Dirac delta function.
\begin{align}
   P_{\rm Poisson}(\hat\delta_R|\delta_R,{\bf x_g})\simeq \delta_{D}\left(\hat\delta_R-\delta_R-\frac{1}{\bar{N}}\right)
\end{align}
Therefore, combining all the terms, we get
\begin{align}
     P(N|{\bf x_g})&\equiv P_\mathrm{TC}(\hat\delta_R) \nonumber\\
     &=\frac{\int d\delta_R\;
       \delta_D\!\left(\hat\delta_R-\delta_R-\frac{1}{\bar N}\right)\,\left(1+\frac{\bar{\xi}(R)}{\sigma^2(R)}\,\delta_R\right)P\left(\delta_R \right)}{ \int \left(1+\frac{\bar{\xi}(R)}{\sigma^2(R)}\,\delta_R \right)P\left(\delta_R \right) d\delta_R }\nonumber\\
    &=\left(1+\frac{\bar{\xi}(R)}{\sigma^2(R)}\,\left(\hat\delta_R-\frac{1}{\bar N}\right)\right)P\left(\hat \delta_R-\frac{1}{\bar N}\right).
\end{align}
The mean value of $\hat\delta_R$ for the tracer-centred PDF is,
\begin{align}
   \langle\hat\delta_R\rangle_\mathrm{TC}&= \int(\hat\delta_R )P_\mathrm{TC}(\hat \delta_R)\,d\hat\delta_R\nonumber\\&=\int\left(\delta_R+\frac{1}{\bar N}\right)\left(1+\frac{\bar{\xi}(R)}{\sigma^2(R)}\,\delta_R\right)P\left( \delta_R\right)\,d\delta_R\nonumber\\&=\frac{1}{\bar N}+\frac{\bar{\xi}(R)}{\sigma^2(R)}\,\int \delta_R^2\,P\left( \delta_R\right)\,d\delta_R = b_g^2\,\bar{\xi}(R)+\frac{1}{\bar N}.
\end{align}
So, the mean of the PDF of the tracer-centred density contrast is non-zero. It is positive, and its amplitude depends on the clustering amplitude of the tracers.
The value is easily understood.  Since $N=\bar{N}\,(1+\hat\delta_R)$, the mean count is 1 (for the particle in the centre) plus the contribution from the mean number of other random particles within $R$, $\bar N$, plus the mean number of correlated pair particles within $R$:  $\bar{N}\,\bar\xi(<R)$.

\section{Supplementary figures}
In this appendix, we collect supplementary figures referred to in the main text.
\label{Apendix}
\begin{figure*}
\centering
\includegraphics[width=0.9\textwidth]{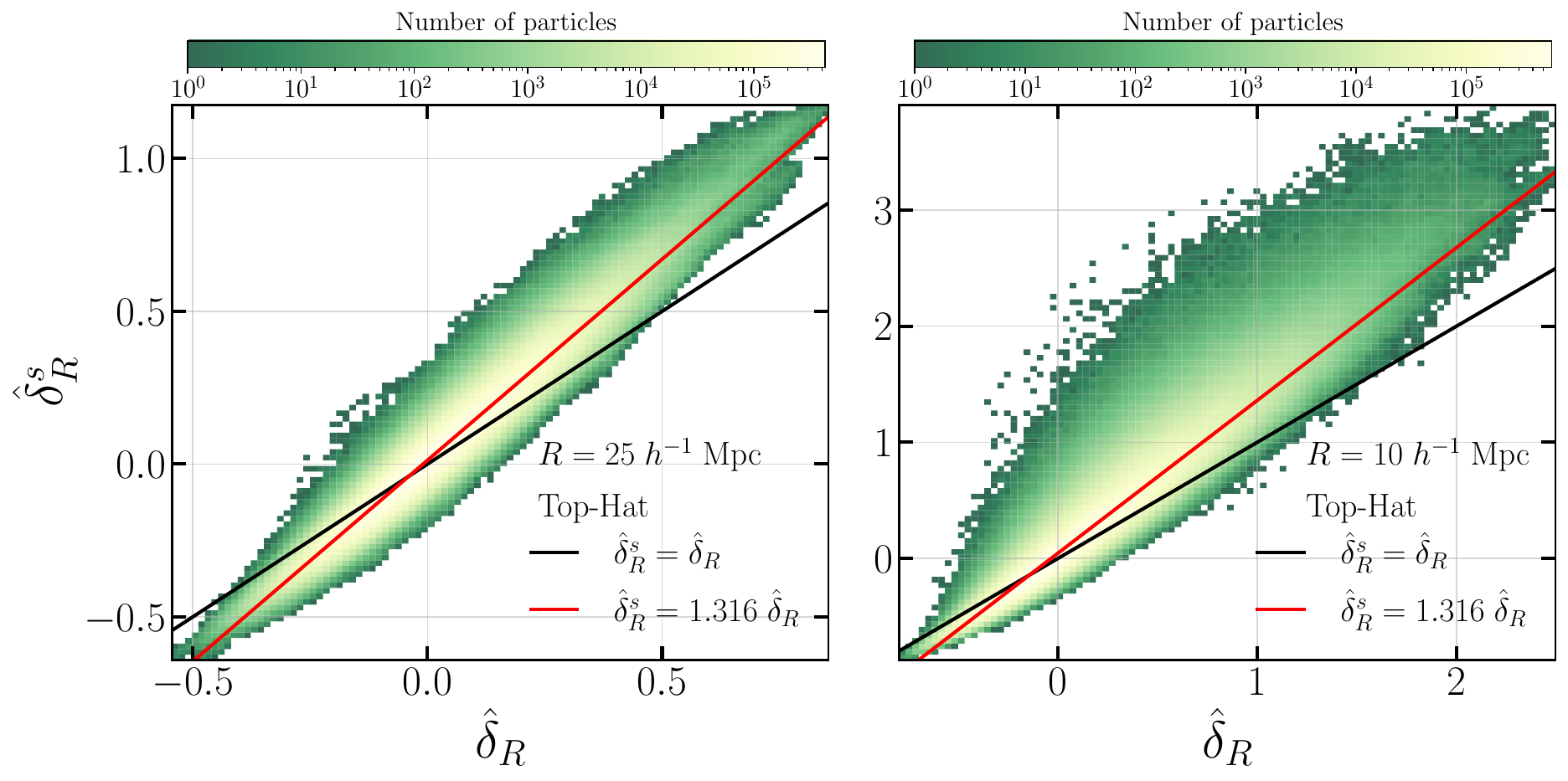}
\caption{Left: A 2D histogram showing the relation between the real-space density and redshift-space density for the mock galaxies when the number density is measured using a smoothing scale of $25\,h^{-1}{\rm Mpc}$. Right: The same for a smoothing scale of $10\,h^{-1} {\rm Mpc}$. In both plots, the red lines show a linear fit to the histogram, where we obtain a slope = $(1+\frac{1}{3}\beta_g)\simeq1.316$, which agrees with Eq.~\ref{eq:deltas}. The black lines denote a linear relation between the real and redshift-space density fields with a slope = $1$, for comparison. We see that for a larger smoothing scale, the scatter around the linear fit decreases significantly.}
\label{fig:scatter_plot}
\end{figure*}
\begin{figure*}
\centering
\includegraphics[width=1.\textwidth]{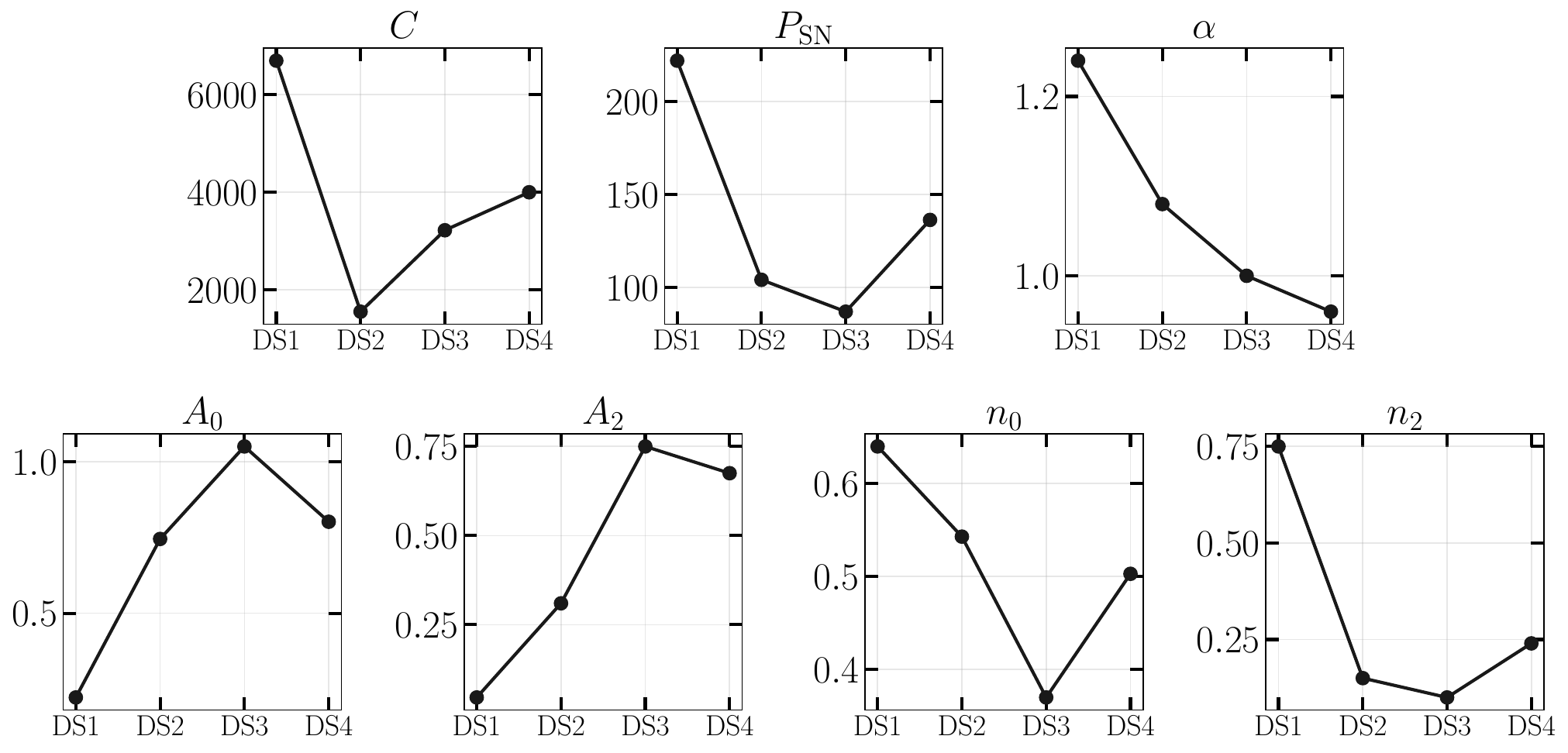}
\caption{The plot shows the values of the 5 nuisance parameters used, along with the value of the $C$ parameter obtained from the constraint equation, i.e. matching the measured variance to that of the model. We also show the shot-noise level for each bin. DS1-4 in each plot indicate density regimes, from low to high. }
\label{fig:nuisance_parameters}
\end{figure*}

\begin{figure*}
\includegraphics[width=0.9\textwidth]{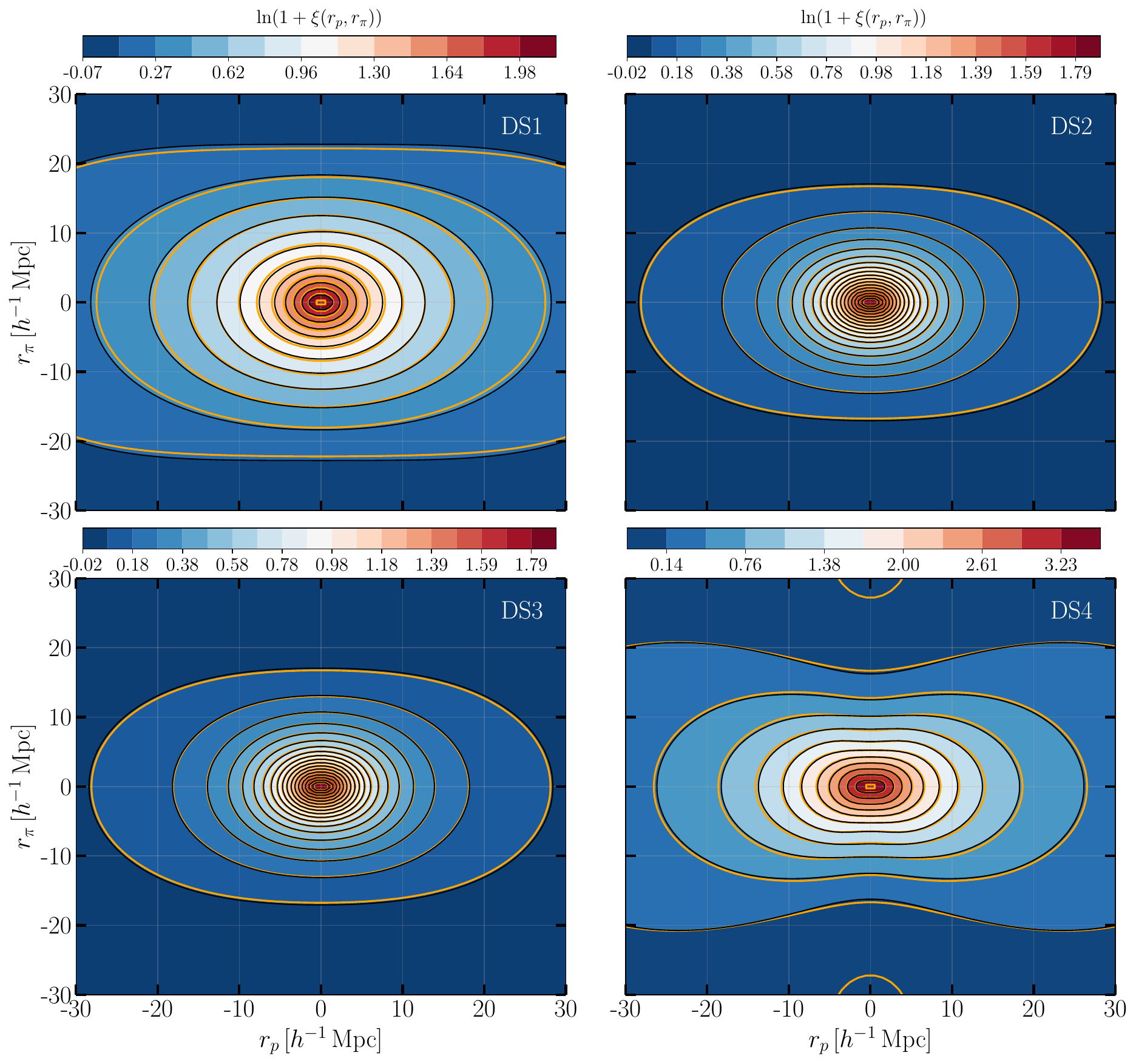}
\caption{These plots show the same 2D correlation functions as shown in Fig.~\ref{fig:2d_correlation_large_scale}, but at smaller scales. These plots highlight the extent of agreement of the model with the measured 2D correlation at small scales.}
\label{fig:2d_correlation_small_scale}
\end{figure*}

\begin{figure*}
\centering
\begin{overpic}[width=0.45\textwidth]{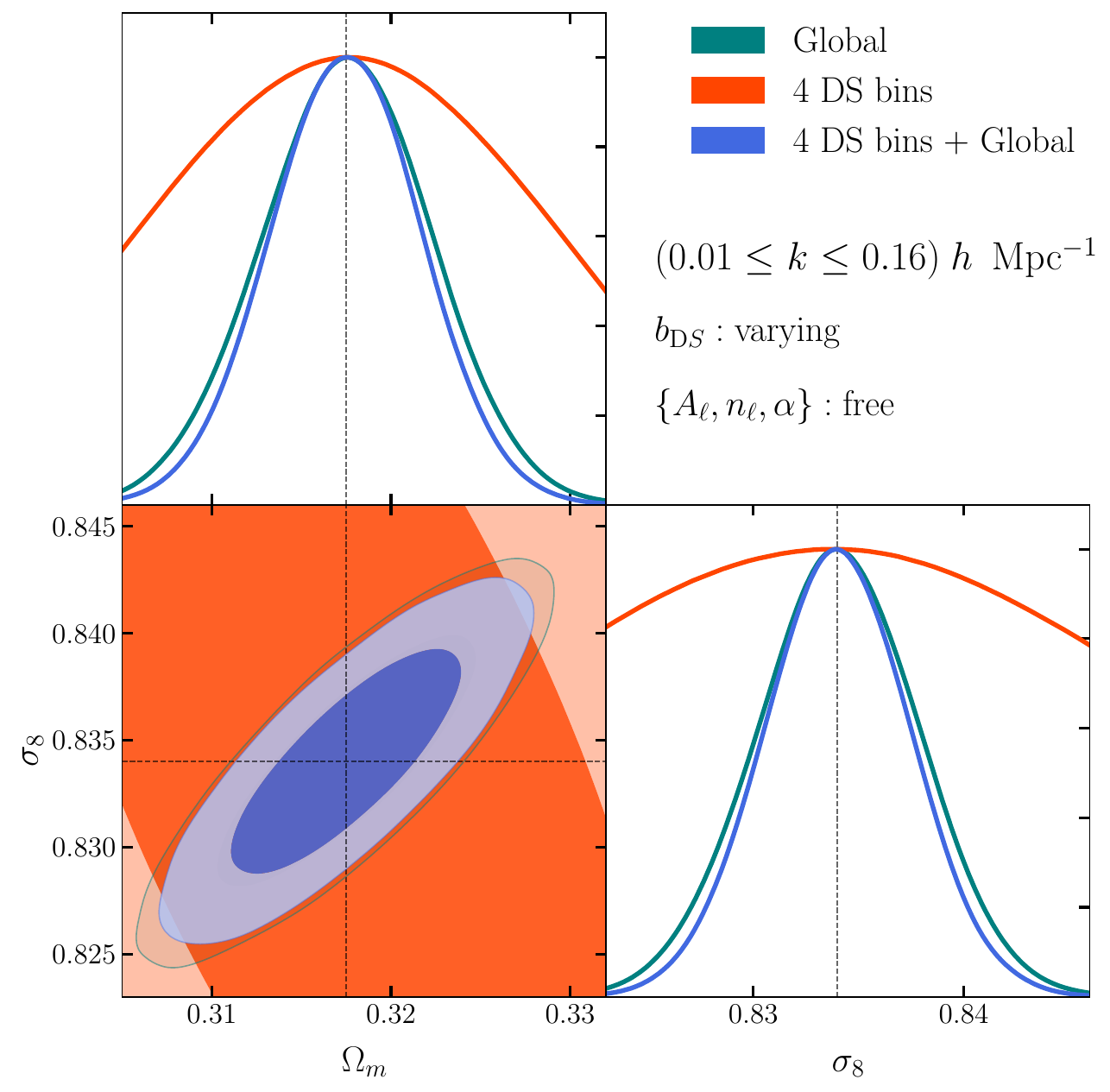}
  \put(13.5,89.5){\text{[A]}}
\end{overpic}%
\begin{overpic}[width=0.45\textwidth]{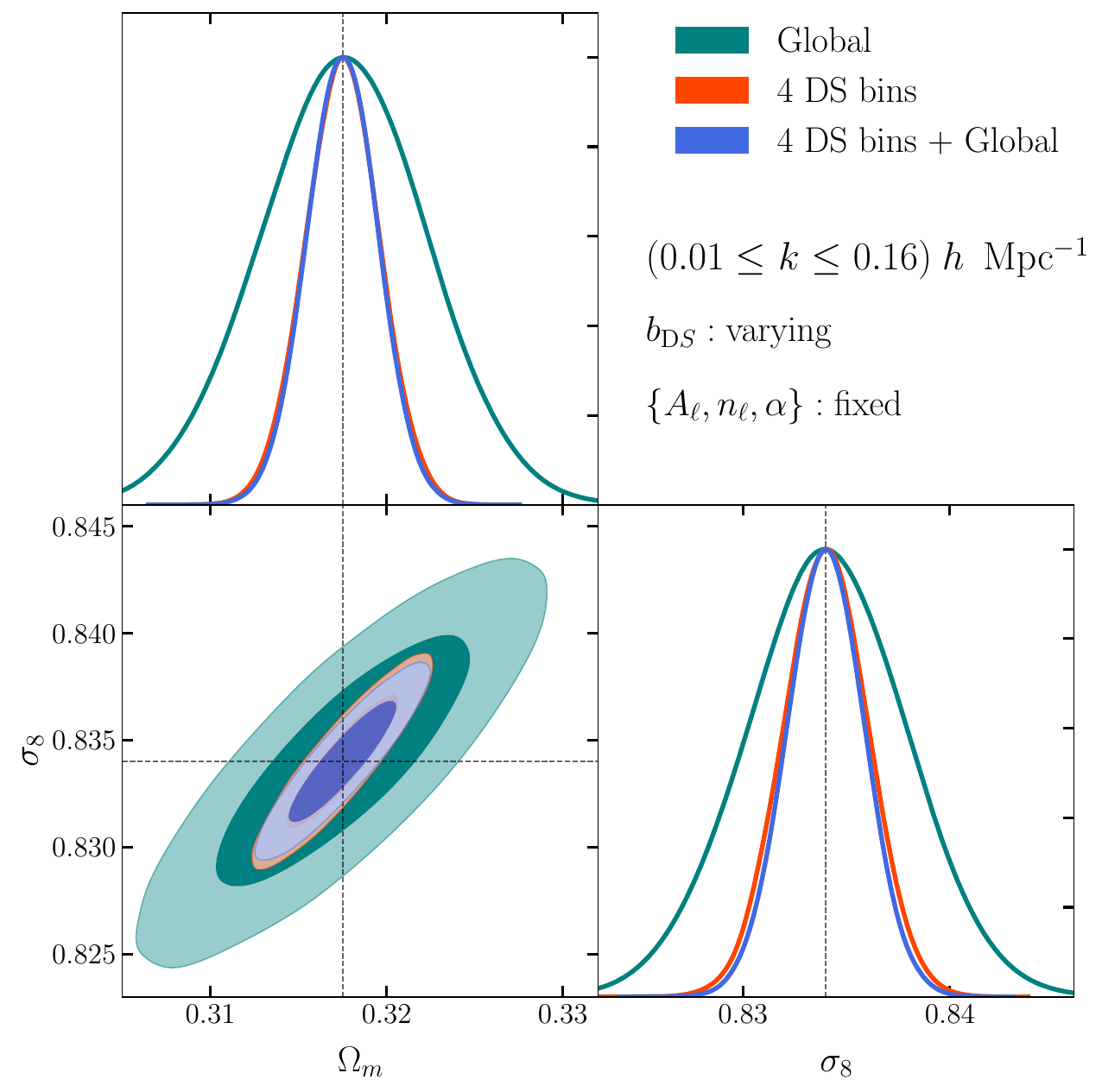}
  \put(13.5,89.5){\text{[B]}}
\end{overpic} \\%
\begin{overpic}[width=0.45\textwidth]{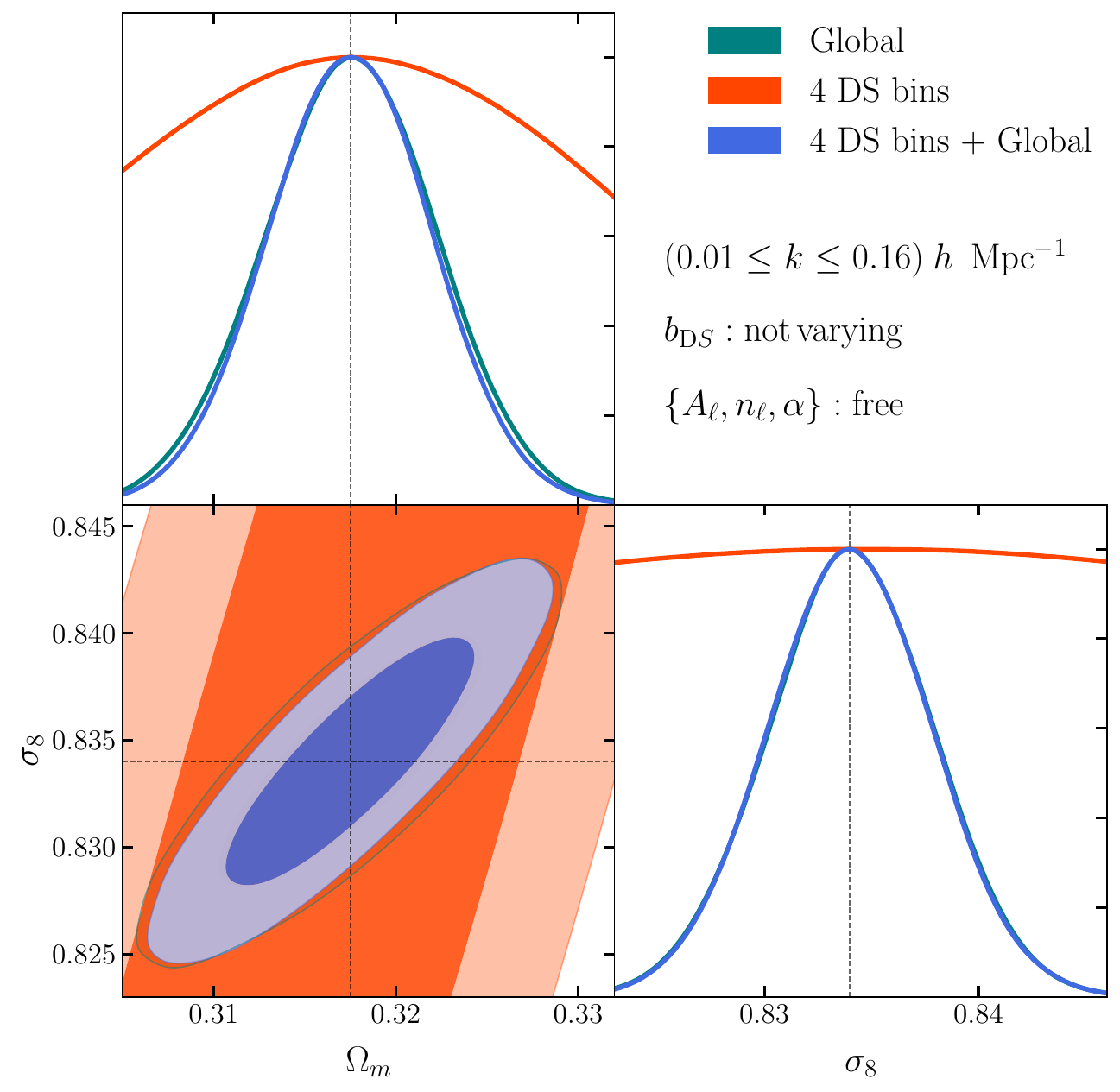}
  \put(13.5,89.5){\text{[C]}}
\end{overpic}%
\begin{overpic}[width=0.45\textwidth]{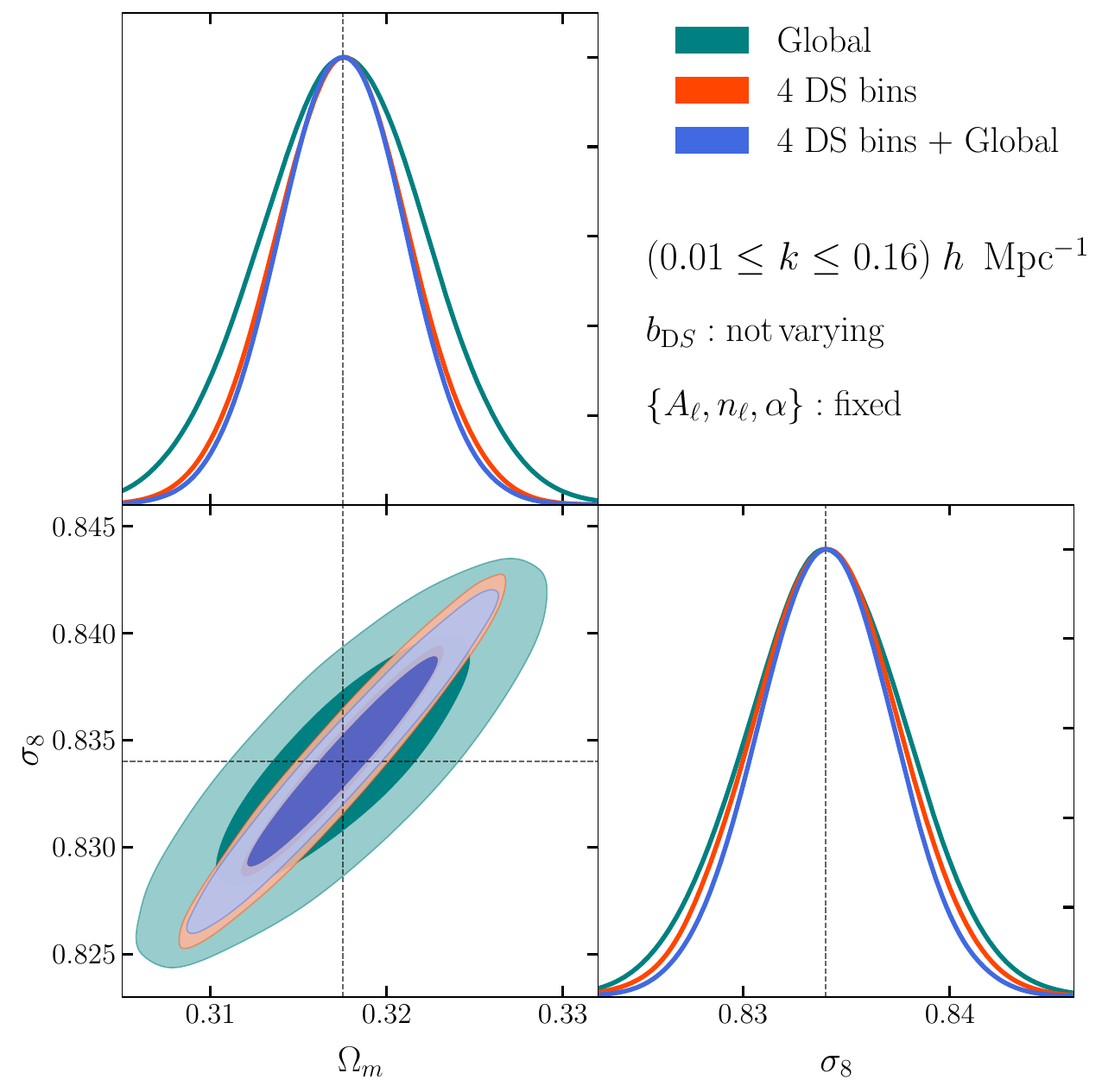}
  \put(13.5,89.5){\text{[D]}}
\end{overpic}

\caption{These four panels are a repetition of the same four earlier cases in Fig.~\ref{fig:Fisher_forecast_mockfactory_diff_scale_cut}, but with the same $k$-cut for the Global power spectra, DS clustering, and the Global+DS combined cases. {\it Top-left}: Constraints on $\sigma_8-\Omega_m$ when marginalising over all the 5 nuisance parameters. In this case, the Global power spectra provide much better constraints than DS clustering. It shows that we lose a lot of constraining power when marginalising over the nuisance parameters, and combining them with the Global case is not enough to make the constraints any better. {\it Top-right}: Constraints when fixing all the nuisance parameters. In this case, DS clustering constrains the cosmological parameters better than the Global power spectra. {\it Bottom-left and bottom-right}: These are repetitions of the {\it top-left and top-right} cases, respectively, but here we have fixed the DS biases $b_{\rm DS}$ to their fiducial values, i.e. the DS biases do not change when we vary the cosmological parameters. We see that fixing the DS biases weakens the constraints from DS cases significantly. These two panels show that in DS clustering, a significant contribution to the total information content comes from the dependence of DS biases on cosmological parameters.}
\label{fig:Fisher_forecast_mockfactory_same_scale_cut}
\end{figure*}
\section{Impact of marginalising over the linear  bias in the Fisher forecast}
\label{app:free_linear_bias}
In the forecasts shown here, we present the conservative case where the observed $P_\ell(k)$ are fixed, and the tracer-centred density split values are also fixed to the same constant fractions of the full sample (e.g. those used in the main text). This makes the linear galaxy bias $b_g$ a free parameter in the model, alongside $\Omega_m$, $\sigma_8$ and the five nuisance parameters presented in our model of DS-clustering. (In practice, we will have some constraints on $b_g$ from e.g., galaxy-lensing correlations.)

\begin{figure*}
\centering
\begin{overpic}[width=0.46\textwidth]{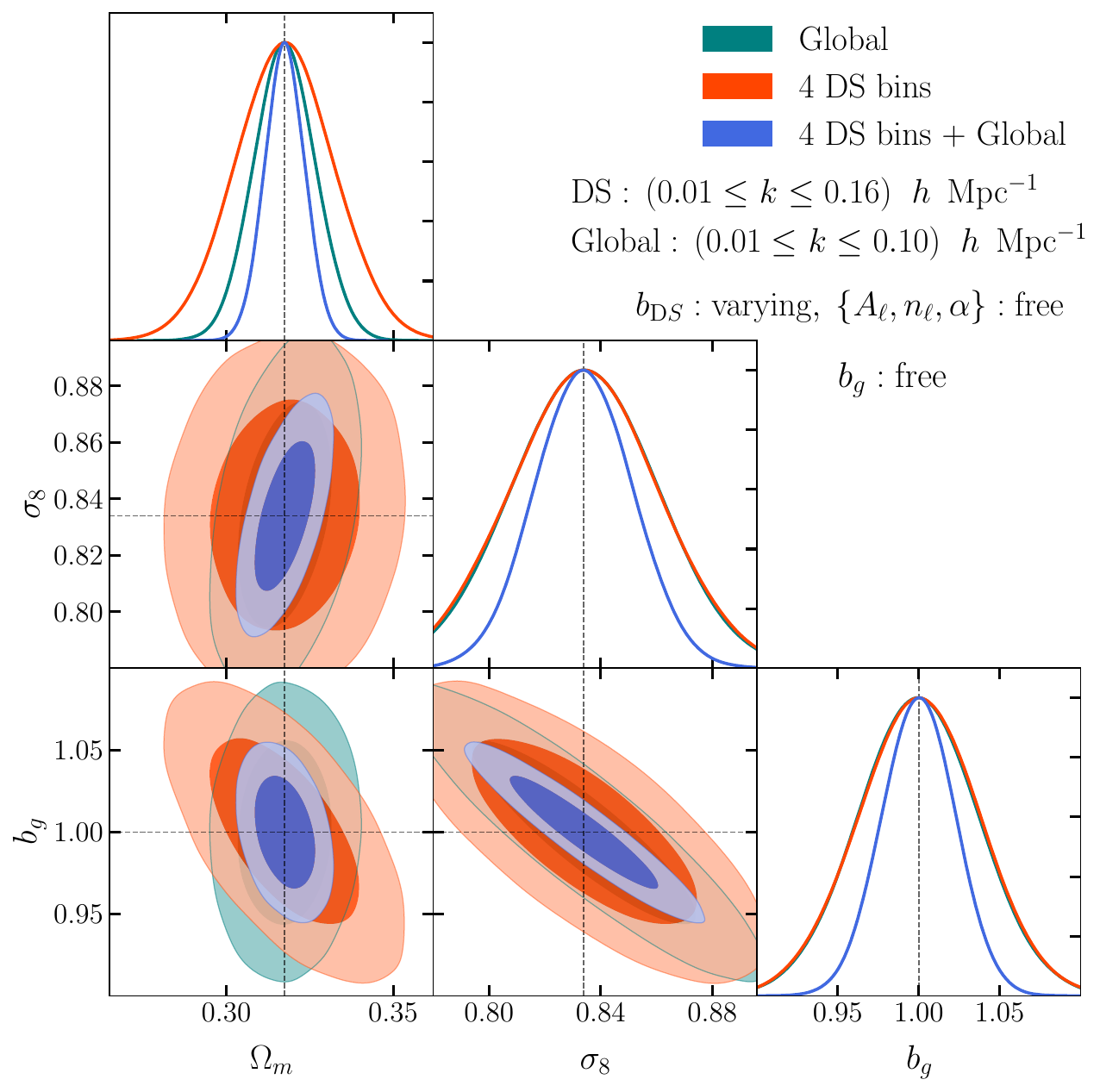}
  \put(13.5,88){\text{[A]}}
\end{overpic}%
\begin{overpic}[width=0.46\textwidth]{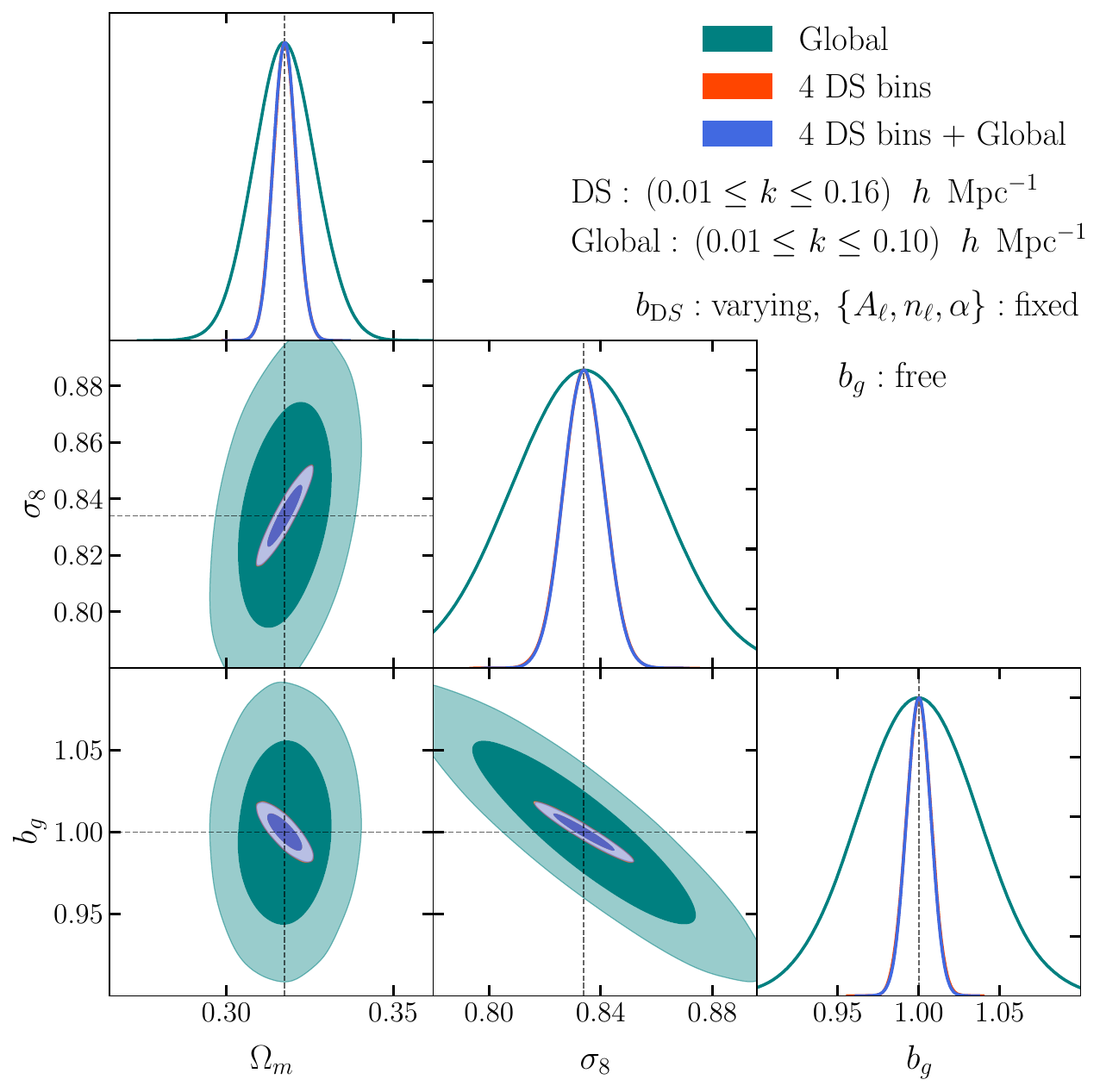}
  \put(13.5,88){\text{[B]}}
\end{overpic} \\%
\begin{overpic}[width=0.46\textwidth]{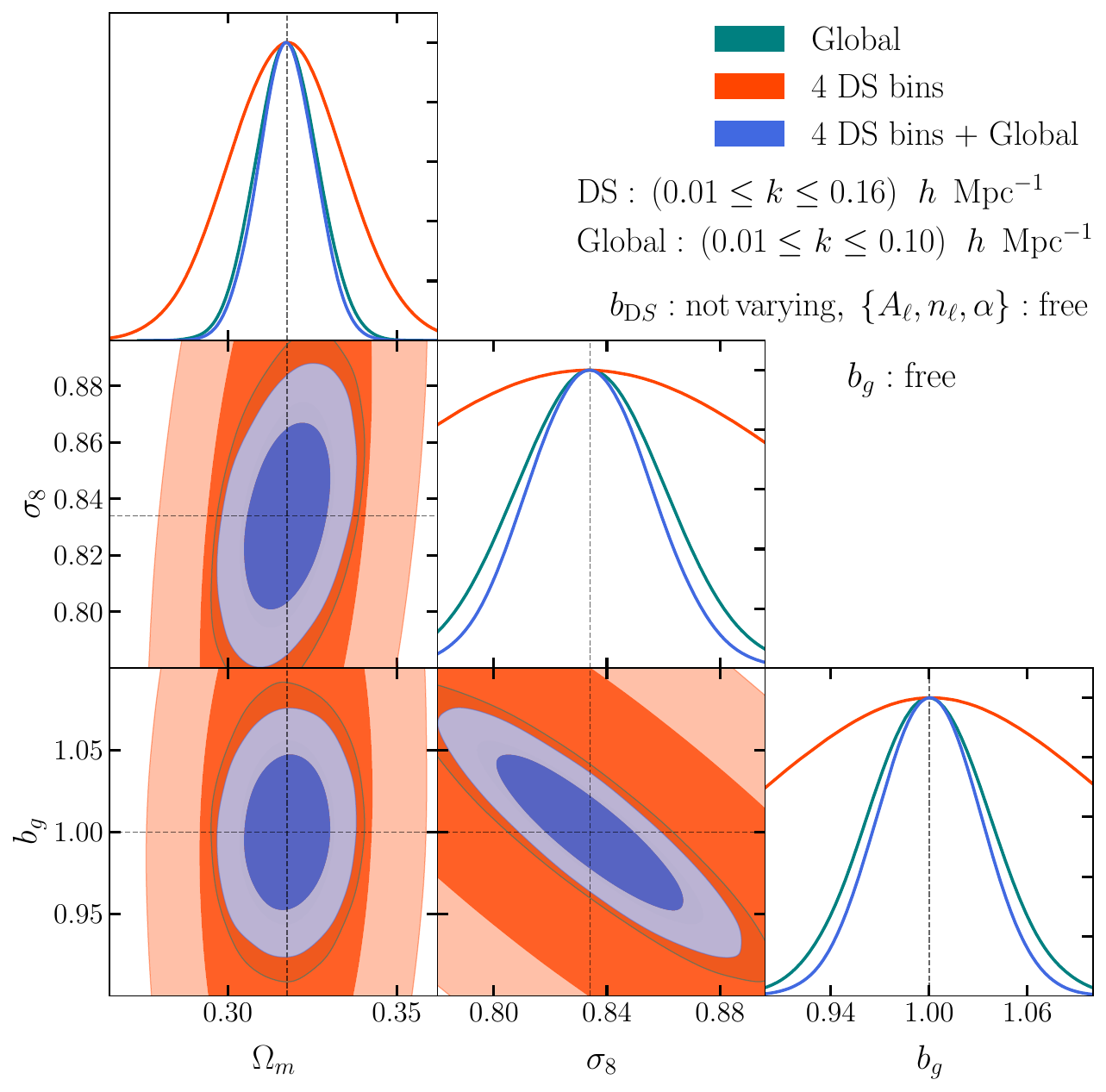}
  \put(13.5,88){\text{[C]}}
\end{overpic}%
\begin{overpic}[width=0.46\textwidth]{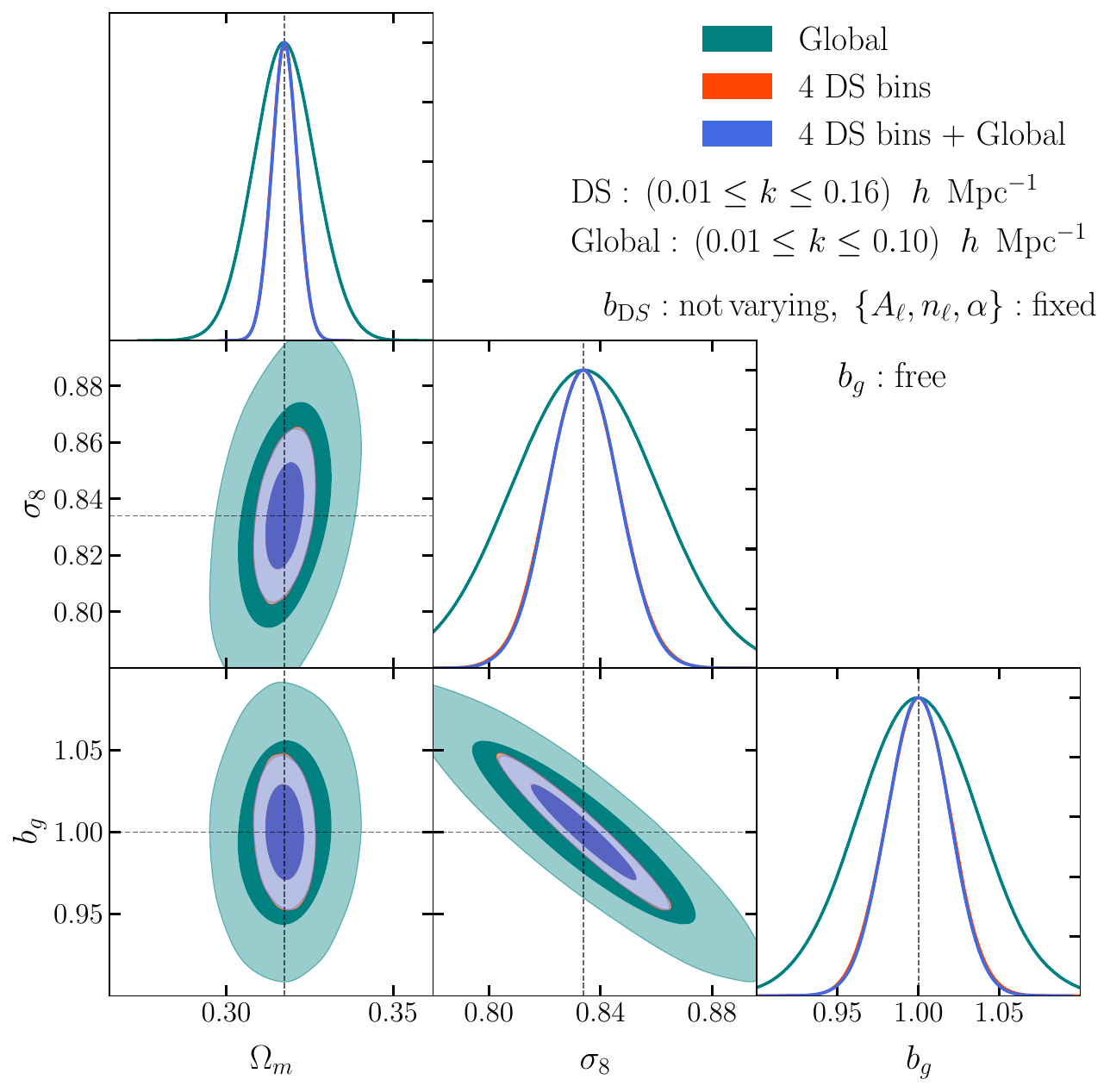}
  \put(13.5,88){\text{[D]}}
\end{overpic}
\caption{Similar to Fig.~\ref{fig:Fisher_forecast_mockfactory_diff_scale_cut}, but with the linear galaxy bias $b_g$ treated as a free parameter. Fisher forecasts for $\sigma_8-\Omega_m$ using the Global redshift-space monopole and quadrupole power spectra (green), the density-split clustering with four density bins (orange), and the combination of the two (blue). {\it Panel A}: the 5 nuisance parameters of the DS model marginalised over. {\it Panel B}: the 5 nuisance parameters fixed. {\it Panels C and D}: repetitions of panels A and B, respectively, with the DS biases $b_{\rm DS}$ fixed to their fiducial values. The $k$-ranges used for the forecasts are shown in the panels.} 
\label{fig:Fisher_forecast_diff_scale_cut_flex_bG}
\end{figure*}

\begin{figure*}
\centering
\begin{overpic}[width=0.45\textwidth]{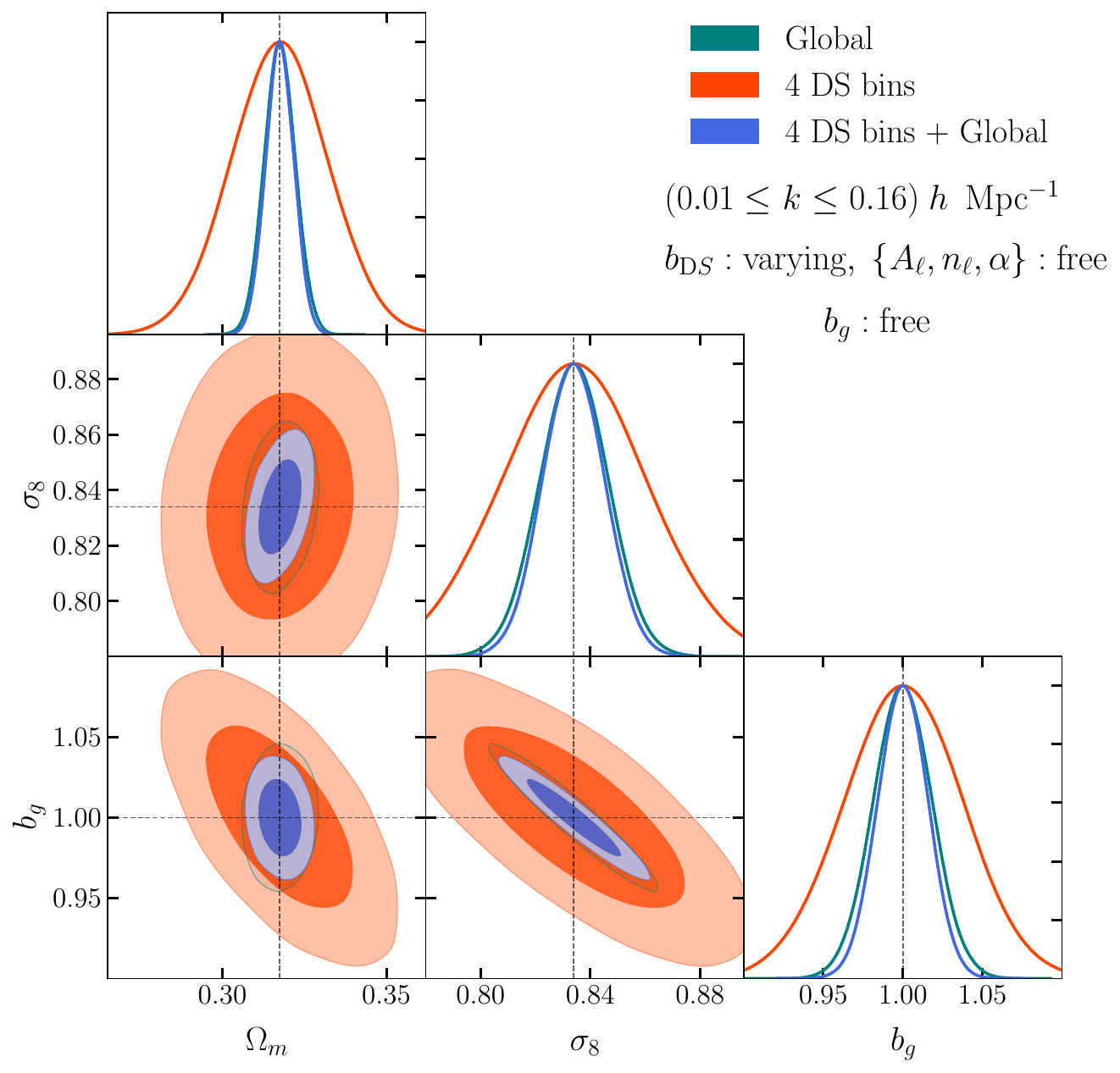}
  \put(13.5,88){\text{[A]}}
\end{overpic}%
\begin{overpic}[width=0.45\textwidth]{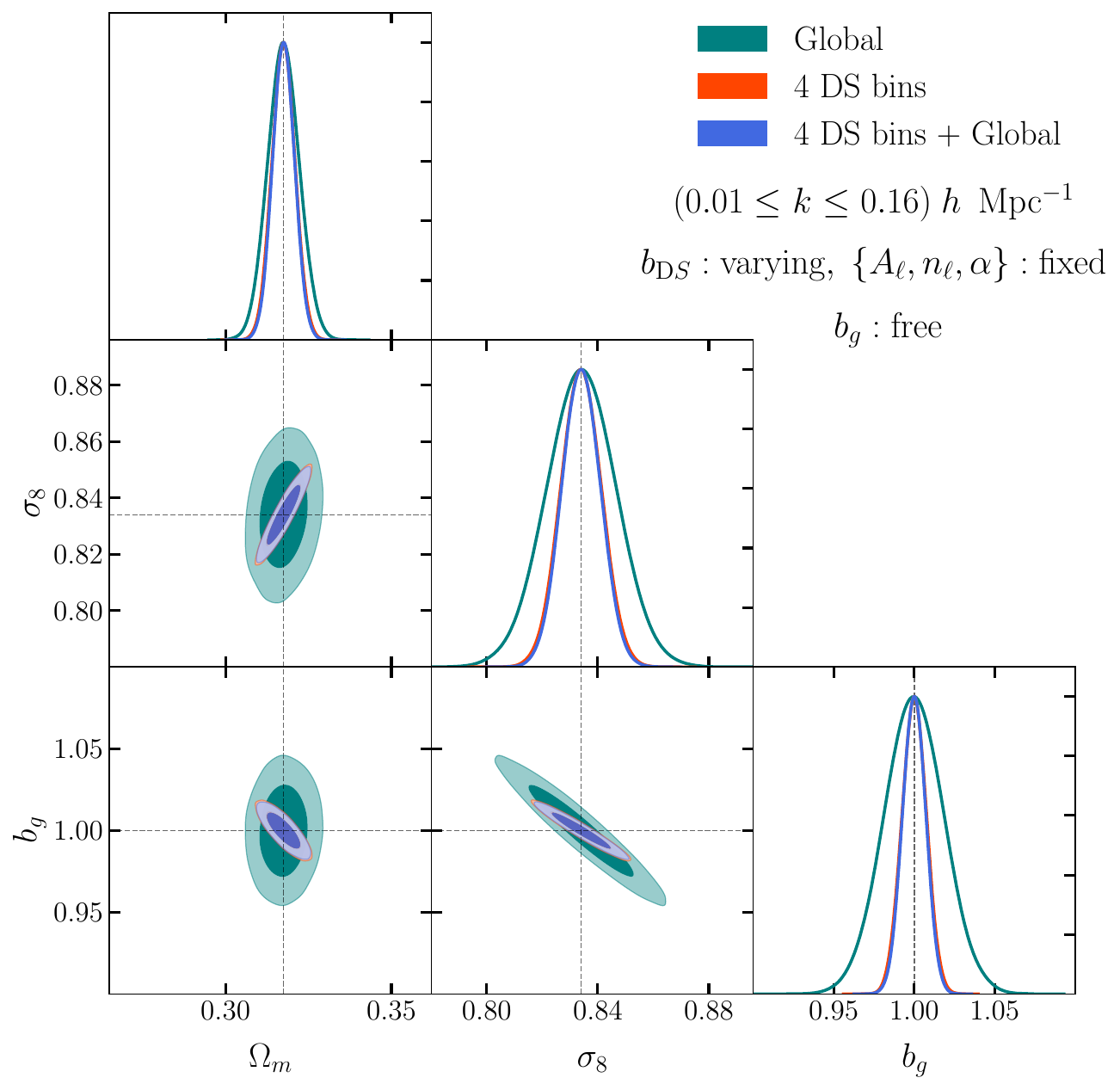}
  \put(13.5,88){\text{[B]}}
\end{overpic} \\%
\begin{overpic}[width=0.45\textwidth]{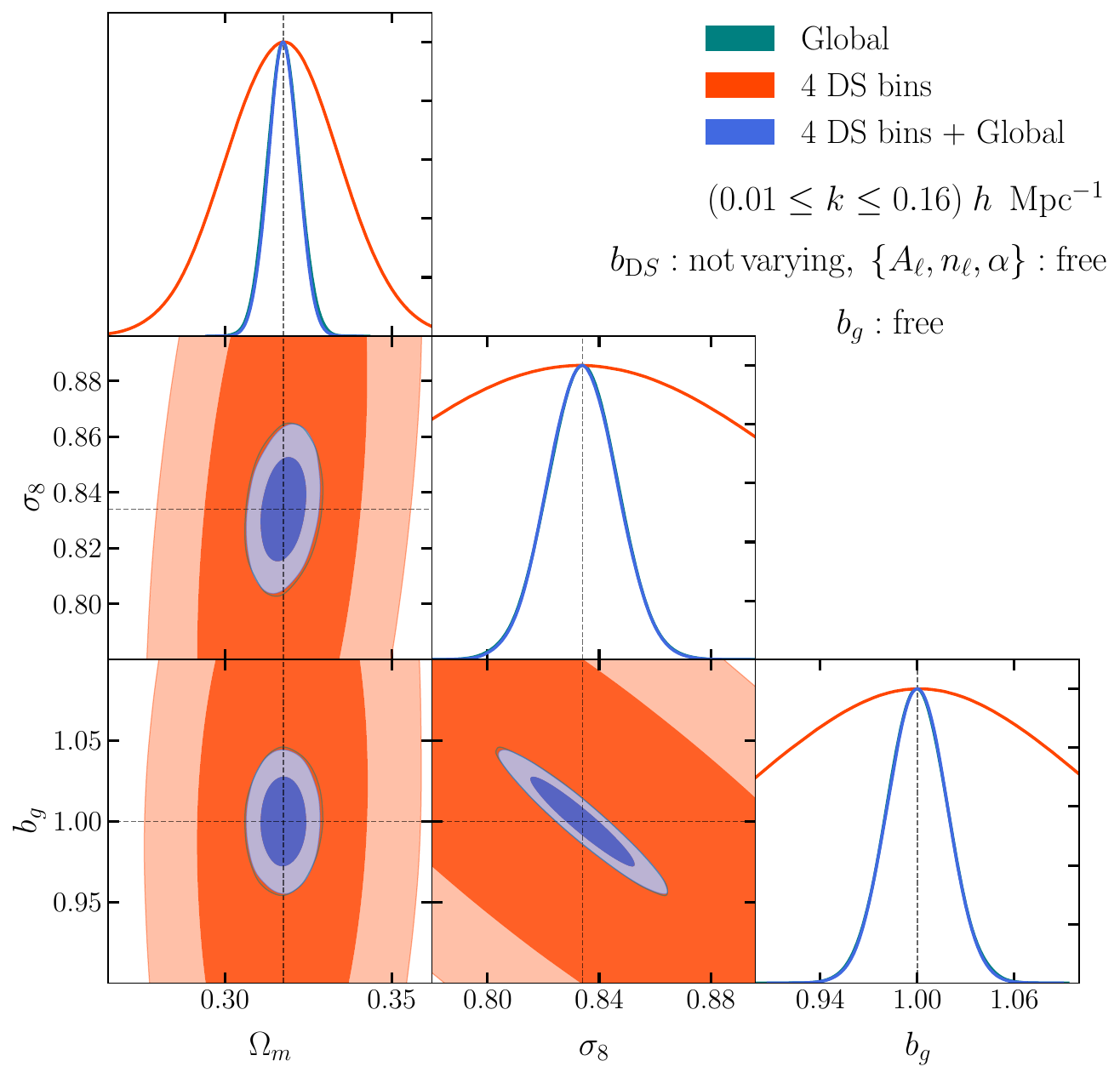}
  \put(13.5,88){\text{[C]}}
\end{overpic}%
\begin{overpic}[width=0.45\textwidth]{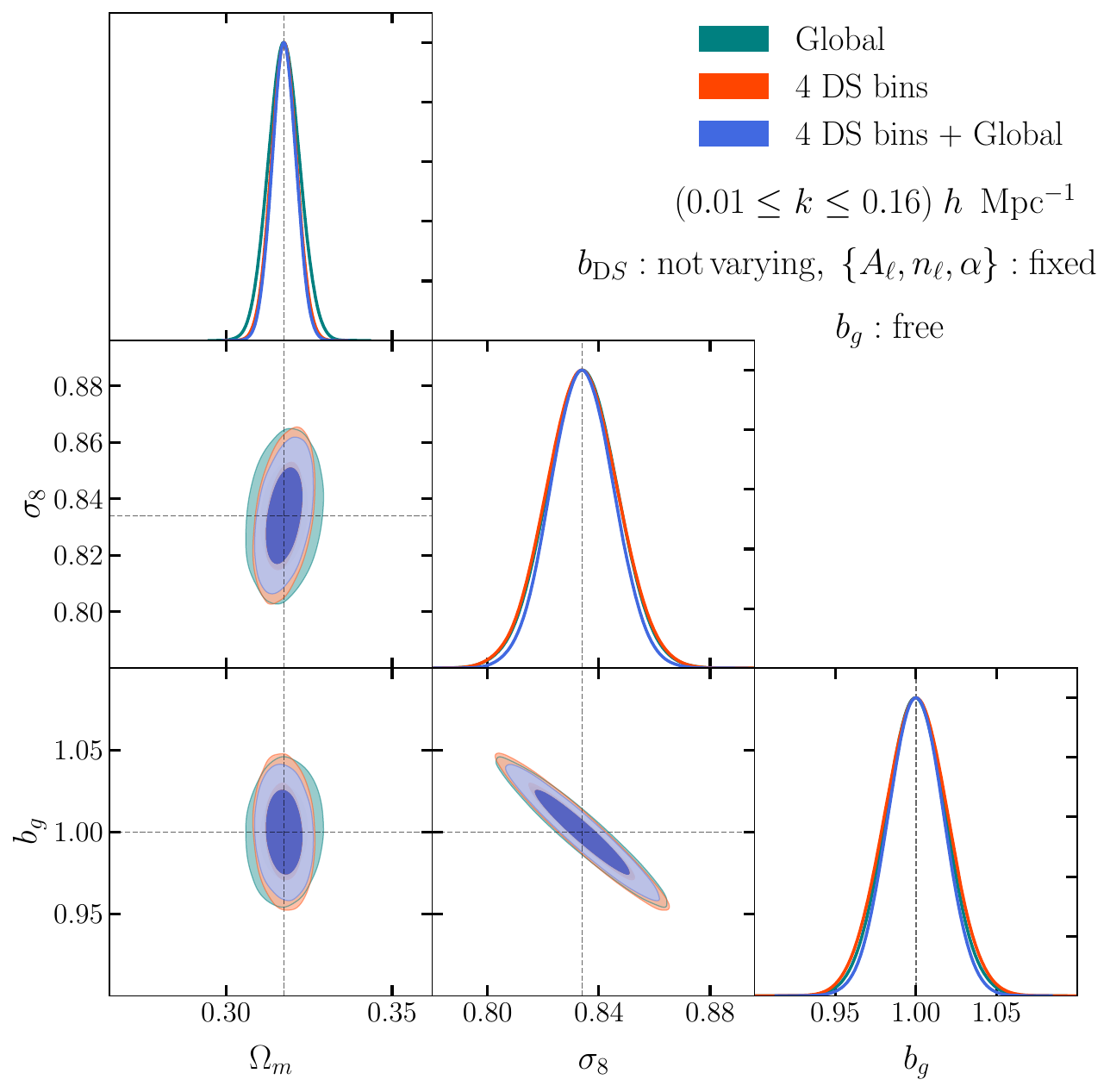}
  \put(13.5,88){\text{[D]}}
\end{overpic}
\caption{Similar to Fig.~\ref{fig:Fisher_forecast_mockfactory_diff_scale_cut}, but the linear galaxy bias $b_g$ is treated as a free parameter. All forecasts use the same $k$-range for the Global power spectra, DS clustering, and the Global+DS combination. {\it Panel A}: the 5 nuisance parameters of the DS model marginalised over. {\it Panel B}: the 5 nuisance parameters fixed. {\it Panels C and D}: repetitions of panels A and B, respectively, with the DS biases $b_{\rm DS}$ fixed to their fiducial values.}
\label{fig:Fisher_forecast_same_scale_cut_flex_bG}
\end{figure*}

The covariance matrices are estimated from the same 2000 {\sc mockfactory} realisations at the fiducial value $b_g=1$, while the derivatives with respect to $b_g$ are computed from the model, by finite differences as in the earlier cases presented in Section~\ref{sec:fisher_forecast}, for both the Global and DS power spectra. For DS clustering, $b_g$ enters through the overall amplitude, the redshift-space factor $\beta_g=f/b_g$, and the DS biases $b_{\rm DS}$, which depend on the variance of the galaxy density field. In the cases where $b_{\rm DS}$ is fixed (bottom panels of Figs.~\ref{fig:Fisher_forecast_diff_scale_cut_flex_bG} and~\ref{fig:Fisher_forecast_same_scale_cut_flex_bG}), it is held at its fiducial value, while we still vary $b_g$ as well as the cosmological parameters. In all cases, $b_g$ is kept free, even for cases where we fix the other nuisance parameters.

The Fisher matrices in Fig.~\ref{fig:Fisher_forecast_diff_scale_cut_flex_bG} and Fig.~\ref{fig:Fisher_forecast_same_scale_cut_flex_bG} show the result of this analysis. We present the constraints on $\Omega_m$, $\sigma_8$, and the linear bias $b_g$. The overall qualitative results remain similar to those in Section~\ref{sec:fisher_forecast}, although, as expected, the constraint on $\sigma_8$ is worse. The galaxy bias plays a role similar to $\sigma_8$. For the Global power spectra, the amplitude depends on the product $b_g\sigma_8$. The degeneracy between $b_g$ and $\sigma_8$ is broken through the RSD effects, i.e. through $\beta_g$, from the ratio of the quadrupole to monopole amplitude, but at $z=2$, the ratio constrains $b_g$, and hence $\sigma_8$ much less precisely than the constraint on $\sigma_8$ when $b_g$ is fixed. Therefore, the overall constraints on $\sigma_8$ become considerably weaker. For the Global case, $\Omega_m$ and $\sigma_8$ remain weakly correlated. At low redshift, marginalising over the linear bias produces an anti-correlation, because $\sigma_8$ is then constrained through $f\sigma_8$ and $f\simeq\Omega_m(z)^{0.55}$ increases with $\Omega_m$. At $z=2$, however, $\Omega_m(z)\simeq0.93$, so $f$ is insensitive to $\Omega_m$, whereas the growth factor, and so $f\sigma_8$ at $z=2$, still decreases with increase in $\Omega_m$, which keeps the degeneracy positive. Since the Global amplitude depends on $b_g\sigma_8$, the same degeneracy is seen in the $\Omega_m-b_g$ plane. This also explains the strong anti-correlation between $\sigma_8$ and $b_g$. 

For the case of DS clustering, there are effects coming from the PDF as well as from the $b_g\sigma_8$ term, which complicates the result. Increasing $b_g$ broadens the galaxy density field, so that, for fixed density thresholds, $b_{\rm DS}$ decreases approximately as $b_g^{-2}$, and the large-scale DS amplitude scales roughly as $(b_g\sigma_8)^{-2}$, rather than as $(b_g\sigma_8)^2$. As a result, the $\Omega_m-b_g$ degeneracy direction flips, similar to the behaviour of the $\Omega_m-\sigma_8$ degeneracy discussed in Section \ref{sec:fisher_forecast}. With $b_g$ free, however, the large-scale DS amplitude constrains only the combination $b_g\sigma_8$, so the anti-correlation between $\Omega_m$ and $\sigma_8$ found with $b_g$ fixed is largely lost, and the $\Omega_m-\sigma_8$ degeneracy becomes mildly positive due to the same $f\sigma_8$ argument. The $\sigma_8-b_g$ degeneracy remains anti-correlated as expected. As in the previous case, fixing $b_{\rm DS}$ significantly reduces the constraining power, highlighting that DS clustering contains extra cosmological information through the DS bias.

It may seem surprising that DS clustering can still constrain $\sigma_8$ and $b_g$ separately, and more tightly than the Global power spectra for the same $k$-range cut-off, when the nuisance parameters are fixed. Indeed, the amplitude alone cannot constrain $\sigma_8$ and $b_g$ separately, but for DS clustering, the $b_{\rm DS}$ parameter depends on where the fixed bin edges are, which is sensitive to both the width and shape of the PDF. While the width of the galaxy PDF changes similarly for both $b_g$ and $\sigma_8$, changing $\sigma_8$ also changes the skewness. Together, they can constrain $b_g$ and $\sigma_8$. 


\end{document}